\documentclass[journal, twoside]{IEEEtran}
\usepackage[T1]{fontenc}

\usepackage{makeidx}
\usepackage{amsmath}
\usepackage{amsfonts}
\usepackage{amssymb}
\usepackage{amsthm}
\usepackage{graphicx}
\usepackage{url}
\usepackage{algorithm}
\usepackage{algorithmic}
\usepackage[caption=false,font=footnotesize]{subfig}

\usepackage{yhmath}

\usepackage{cite}

\usepackage{color}

\newtheorem{definition}{Definition}
\newtheorem{theorem}{Theorem}
\newtheorem{corollary}{Corollary}
\newtheorem{lemma}{Lemma}

\ifCLASSINFOpdf
\else
\fi
\begin{document}
%
\title{Accuracy of Homology based Coverage Hole Detection for Wireless Sensor Networks on Sphere}
%
%
%

\author{Feng Yan, ~\IEEEmembership{Member,~IEEE,}
        Philippe Martins, ~\IEEEmembership{Senior Member,~IEEE,} and
        Laurent Decreusefond
\thanks{F. Yan was with the Network and Computer Science Department, 
TELECOM ParisTech, Paris, France. He is currently with the Networks, Security and 
Multimedia Department, TELECOM Bretagne, Rennes, France. (e-mail: feng.yan@telecom-bretagne.eu) 

P. Martins and L. Decreusefond are with the Network 
and Computer Science Department, TELECOM ParisTech, Paris, France.
(e-mail: martins@telecom-paristech.fr, decreuse@telecom-paristech.fr.)
A part of this paper has been published in IEEE ICC 2012.}}

%
%

\markboth{IEEE TRANSACTIONS ON WIRELESS COMMUNICATIONS}%
{Yan \MakeLowercase{\textit{et al.}}: Accuracy of Homology based Coverage Hole Detection for Wireless Sensor Networks on Sphere}
%



\maketitle

\begin{abstract}
Homology theory has attracted great attention because it can provide novel
and powerful solutions to address coverage problems in wireless sensor networks. 
They usually use an easily computable algebraic object, Rips
complex, to detect coverage holes. But Rips complex may miss some coverage
holes in some cases. In this paper, we investigate homology-based coverage hole
detection for wireless sensor networks on sphere. The case when Rips complex
may miss coverage holes is first identified. Then we choose the proportion
of the area of coverage holes missed by Rips complex as a metric to evaluate
the accuracy of homology-based coverage hole detection approaches. Closed-form 
expressions for lower and upper bounds of the accuracy
are derived. Asymptotic lower and upper bounds are also investigated when the
radius of sphere tends to infinity. Simulation results are well consistent with the analytical
lower and upper bounds, with maximum differences of 0.5\% and 3\% respectively.
Furthermore, it is shown that the radius of sphere has little impact
on the accuracy if it is much larger than communication and sensing
radii of each sensor.
\end{abstract}


%
\IEEEpeerreviewmaketitle

\section{Introduction}
%
%
%
%
\IEEEPARstart{W}{ireless} sensor networks (WSNs) have attracted considerable
research attention due to their large number of potential applications such as
battlefield surveillance, environmental monitoring and intrusion
detection. Many of these applications require a reliable detection of
specified events. Such requirement can be guaranteed only if the
target field monitored by a WSN contains no coverage holes, that is to
say regions of the domain not monitored by any sensor. But coverage holes
can be formed for many reasons, such as random deployment, energy
depletion or destruction of sensors. Consequently, it is essential to
detect and localize coverage holes in order to ensure the full
operability of a WSN.

Most existing works on coverage hole issues mainly focus on 
two-dimensional (2D) plane or three-dimensional (3D) full space.
There is few work on 3D surfaces. But in some real applications, such as 
volcano monitoring \cite{WLW06} and forest monitoring \cite{MHL09},
the target fields are complex surfaces. So it is also important to consider
the coverage hole detection problem of WSNs on surfaces. On the other hand, 
from theoretical point of view, the coverage on 3D surfaces is quite a 
different problem from its counterpart in 2D plane or
3D full space. As sphere is the simplest case of 3D surfaces, we choose it
as the first step for the analysis in this paper, like the authors 
did in \cite{GK00} for throughput capacity analysis.

There are already extensive works on the coverage hole detection problem for WSNs in 
2D plane and 3D space. Some of these works used either precise information 
about sensor locations \cite{FGG04, WCL04, ZZF09, HT03, HTL04} or accurate 
relative distances between neighbouring sensors \cite{B08, B12} to detect 
coverage holes. The requirement of precise location or distance information
substantially limits their applicability since acquiring such information
is either expensive or impractical in many settings. Thus connectivity-based
approaches are of great interest for us. In this category, homology-based
schemes have received special attention because of its powerfulness for
coverage hole problems in WSNs. 

Homology theory was first adopted by Ghrist and his collaborators in \cite{DSG05, DSG07, GM05}
to address the coverage problems in WSNs. They introduced a combinatorial object,
$\check{\textrm{C}}$ech complex, which uses sensing ranges of nodes to fully characterize coverage properties of a WSN (existence and locations of holes).
Unfortunately, the construction of this object is of very high 
complexity \cite{CO08} even if
the precise location information about sensors is provided. 
Thus, they introduced another more easily computable complex,
Vietoris-Rips complex (we will abbreviate the term to Rips complex in this paper). 
This complex is constructed with the sole
knowledge of the connectivity graph of the network and gives an
approximate coverage by simple algebraic calculations. 
Considering the ease of Rips complex construction, some homology-based 
algorithms were proposed in \cite{ME06, MJ07, TJ10} 
to use Rips complex to detect coverage holes. But all these homology-based
approaches do not consider the cases that
Rips complex may miss some special coverage holes. If the proportion
of the area of coverage holes missed by Rips complex is low enough,
then it is acceptable to use these methods for coverage hole detection.
If the proportion is too high to be unacceptable, then it may not be proper
to use these methods. Therefore, in order to evaluate the accuracy of homology-based 
coverage hole detection approaches, it is of paramount importance to 
analyse the coverage holes missed by Rips complex.

The main contributions of our paper are as follows. First, the relationship between
$\check{\textrm{C}}$ech complex and Rips complex in terms of coverage
hole on sphere is analysed. Furthermore, the case that Rips complex
may miss coverage holes is identified and it is found that a hole in
a $\check{\textrm{C}}$ech complex missed by a Rips complex must be 
bounded by a spherical triangle. Based on that, a formal definition
of spherical triangular hole is given.

Second, the proportion of the area of spherical triangular holes is chosen
as a metric to evaluate the accuracy of homology-based coverage hole 
detection. Such proportion is analysed under a homogeneous setting and it
is related to the communication and sensing radii of each sensor. 
Closed-form expressions for lower and upper bounds of the proportion 
are derived. Asymptotic lower and upper bounds are also investigated
when the radius of sphere tends to infinity.

Third, extensive simulations are performed to evaluate impacts of
communication and sensing radii, radius of sphere on proportion
of the area of spherical triangular holes. It is shown that simulation 
results are well consistent with the analytical lower bound, with a maximum 
difference of 0.5\%, and consistent with the analytical upper bound, with
a maximum difference of 3\%. Furthermore, simulation results
show that the radius of sphere has little impact on the proportion when
it is much larger than communication and sensing radii.   

The rest of the paper is organised as follows. Section II presents the related work.
In Section III, the network model and the formal definition of spherical
triangular hole are given. Closed-form lower and upper bounds for proportion
of the area of spherical triangular holes are derived
in Section IV. Section V compares simulation results and analytical bounds.
Finally, Section VI concludes the paper.

\section{Related work}

Since this paper aims to evaluate the ratio of the area of coverage holes missed
by homology-based approaches, we present the related work in terms of two aspects: 
coverage hole detection approaches and analytical coverage ratio evaluation. 

\subsection{Coverage hole detection approaches}

Many approaches have been proposed for coverage hole detection in WSNs.
They can be generally classified into three categories: location-based, 
range-based and connectivity-based. 

Location-based approaches are usually
based on computational geometry with
tools such as Voronoi diagram and Delaunay triangulations, to discover
coverage holes \cite{FGG04, WCL04, ZZF09}. Range-based approaches
attempt to discover coverage holes by using only relative
distances between neighbouring sensors \cite{B08, B12}.
These two types of approaches need either precise location information 
or accurate distance information, which restricts their applications since
such information is not easy to obtain in many settings. 

In connectivity-based approaches, homology-based schemes attract particular
attention due to its powerfulness for coverage hole detection.
De Silva \emph{et al.} first proposed a centralized 
algorithm that detects coverage hole via homology in \cite{DSG07}. They 
constructed the Rips complex corresponding to the communication graph of the 
network and determined the coverage by verifying whether the first homology
group of the Rips complex is trivial. Then the above ideas were first implemented
in a distributed way in \cite{ME06}. It is shown that combinatorial Laplacians are the 
right tools for distributed computation of homology groups and can be used 
for decentralized coverage verification. In \cite{MJ07}, a gossip-like 
decentralized algorithm for computation of homology groups was proposed. 
In \cite{TJ10}, a decentralized scheme based on Laplacian flows was 
proposed to compute a generator of the first homology group. All these homology-based
algorithms may be also used to detect coverage holes for WSNs on surfaces, but they 
do not consider the cases that Rips complex
may miss some special coverage holes. One of our objectives in this paper
is to identify such cases.

\subsection{Analytical coverage ratio evaluation} 

Extensive research has been done to analyse coverage ratio of a WSN in 2D plane or on 3D surfaces. 
In \cite{LT04}, the authors studied the coverage properties of large-scale
sensor networks and obtained the fraction of the area covered by sensors. 
The sensors are assumed to have the same sensing range and 
are distributed according to a homogeneous Poisson point process (PPP) in plane.
In \cite{WY06}, the authors studied how the 
probability of $k$-coverage changes with the sensing radius 
or the number of sensors, given that sensors are deployed
as either a PPP or a uniform point process. 
In addition, the distance distribution between two points
in random networks was derived in \cite{M12}. Their results can be 
used to derive the fraction of areas covered by at least $k$-sensors.
All the above studies only considered homogeneous cases. 
In \cite{LP06}, the coverage problem in planar heterogeneous sensor 
networks are investigated and analytical expressions of coverage
are derived. Their formulation is more general in the sense that
sensor can be deployed according to an arbitrary stochastic distribution,
or can have different sensing capabilities or can have arbitrary
sensing shapes. Based on their results, the authors in \cite{ZLW09}
derived the expected coverage ratio of sensors under stochastic 
deployment on 3D surface. Similarly, the expected coverage ratio 
under stochastic deployment on 3D rolling terrains was derived in 
\cite{LM12}. In \cite{LHZ12}, a point in a plane
is defined to be tri-covered if it lies inside a triangle formed
by three nodes, and the probability of tri-coverage was analysed.

All the above research considered only coverage ratio
problems, without considering coverage hole detection issues. 
Their analysis is thus not specific to any coverage hole detection
approaches. We provided some initial 
results about the proportion of the area of triangular holes for 
WSNs in 2D plane in \cite{YMD12}. In this paper, we aim to analyse the proportion of 
the area of coverage holes missed by homology-based coverage hole 
detection approaches for WSNs on sphere and compare it with the case
in 2D plane. 

\section{Models and definitions} \label{secmod}

Consider a collection of stationary sensors (also called nodes)
on a sphere $\mathbb{S}^2$ with radius $R$. The sensors are deployed
according to a homogeneous PPP with intensity $\lambda$.
For any two points $p_1$ and $p_2$ on $\mathbb{S}^2$,
the distance between them $d(p_1, p_2)$ is defined to be the 
great circle distance, which is the shortest distance between  
them measured along a path on the surface of the sphere. 
As usual, isotropic radio propagation is assumed. All sensors 
have the same sensing radius $R_s$ and communication radius $R_c$
on $\mathbb{S}^2$. It means for any sensor located at $v$ on $\mathbb{S}^2$,
any point $p$ on $\mathbb{S}^2$ with $d(v,p) \leq R_s$ is inside the sensing range of the sensor;
and for any two sensors located at $v_i, v_j$ on $\mathbb{S}^2$, they can communicate
with each other if $d(v_i,v_j) \leq R_c$.  In addition, we assume
$R_s \ll R$, $R_c \ll R$.

\begin{figure}[ht]
  \centering
  \includegraphics[width = 0.4\textwidth]{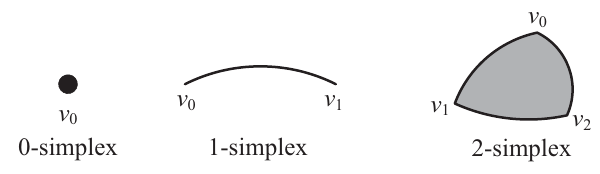}
  \caption{0-, 1- and 2-simplex}
  \label{simplex}
\end{figure}

Before defining the two combinatorial objects, known as 
$\check{\textrm{C}}$ech complex and Rips complex, 
it is necessary to give a brief introduction to some tools used
in the paper. For further readings, see \cite{ARM83, mun84, HAT02}.
Given a set of points $V$, a \textit{k}-simplex is an
unordered set $[v_0, v_1, ..., v_k] \subseteq V$ where $v_i \neq
v_j$ for all $i \neq j$, $k$ is the dimension of this simplex. The faces of this \textit{k}-simplex consist of
all (\textit{k}-1)-simplex of the form $[v_0, ..., v_{i-1},v_{i+1},
..., v_k]$ for $0 \leq i \leq k$. For example, on a sphere $\mathbb{S}^2$,
a 0-simplex $[v_0]$ is a vertex , a 1-simplex $[v_0,v_1]$ is the shorter arc
of the great circle passing through $v_0$ and $v_1$, a 2-simplex 
$[v_0, v_1, v_2]$ is a spherical triangle $v_0v_1v_2$ with its interior 
included, see Figure \ref{simplex}. An abstract simplicial complex is a collection of simplices which 
is closed with respect to inclusion of faces. A $k$-dimensional abstract simplical complex $\mathcal{K}$
is an abstract simplicial complex where the largest dimension of any simplex in $\mathcal{K}$ is $k$.

Let $\mathcal{V}$ denote the set of
sensor locations in a WSN on $\mathbb{S}^2$ with radius $R$ and $\mathcal{S}=\{s_v,\, v\in \mathcal{V}\}$  denote the collection of sensing
ranges of these sensors: for a location $v$,  $s_v = \{ x \in\mathbb{S}^2: d(x,v) \leq R_s\}$. 
Then $\check{\textrm{C}}$ech complex and Rips complex can be defined 
as follows \cite{DSG05,DSG07}.

\begin{definition}[$\check{\textrm{C}}$ech complex] 
Given a finite collection of sensing ranges $\{s_v,\, v\in \mathcal{V}\}$,
the $\check{\textrm{C}}$ech complex of the collection, $\check{\textrm{C}}(\mathcal{V})$, 
is the abstract simplicial complex whose \textit{k}-simplices correspond to non-empty
intersections of k + \emph{1} distinct elements of $\{s_v,\, v\in \mathcal{V}\}$.
\end{definition}

\begin{definition} [Rips complex]
Given a finite set of points $\mathcal{V}$ on 
$\mathbb{S}^2$ and a fixed radius $\epsilon$, the Rips complex of $\mathcal{V}$,
$\mathcal{R}_\epsilon$($\mathcal{V}$), is the abstract simplicial complex
whose \textit{k}-simplices correspond to unordered \emph{(\textit{k} +1)}-tuples of 
points in $\mathcal{V}$ which are pairwise within distance $\epsilon$ of each other.
\end{definition}

According to the definitions, the
$\check{\textrm{C}}$ech complex and Rips complex of the WSN, respectively denoted by $\check{\textrm{C}}_{R_s}
(\mathcal{V})$ and $\mathcal{R}_{R_c}(\mathcal{V})$, can be
constructed as follows: a \textit{k}-simplex $[v_0, v_1,\cdots,v_k]$ belongs to $\check{\textrm{C}}_{R_s}
(\mathcal{V})$ whenever $\cap_{l=0}^k s_{v_l}\not =
\emptyset$ and a \textit{k}-simplex $[v_0, v_1,\cdots,v_k]$ belongs to
$\mathcal{R}_{R_c}(\mathcal{V})$ whenever $d(v_l, v_m)\le R_c$ for all
$ 0\le l<m\le k$. In addition, since we consider only coverage holes on 
the sphere $\mathbb{S}^2$, it is sufficient to construct 
2-dimensional $\check{\textrm{C}}$ech complex and 2-dimensional Rips complex of the WSN,
denoted as $\check{\textrm{C}}_{R_s}^{(2)}(\mathcal{V})$ and $\mathcal{R}_{R_c}^{(2)}(\mathcal{V})$
respectively.

\begin{figure}[ht]
  \centering \subfloat[]{\includegraphics[width=0.2\textwidth]{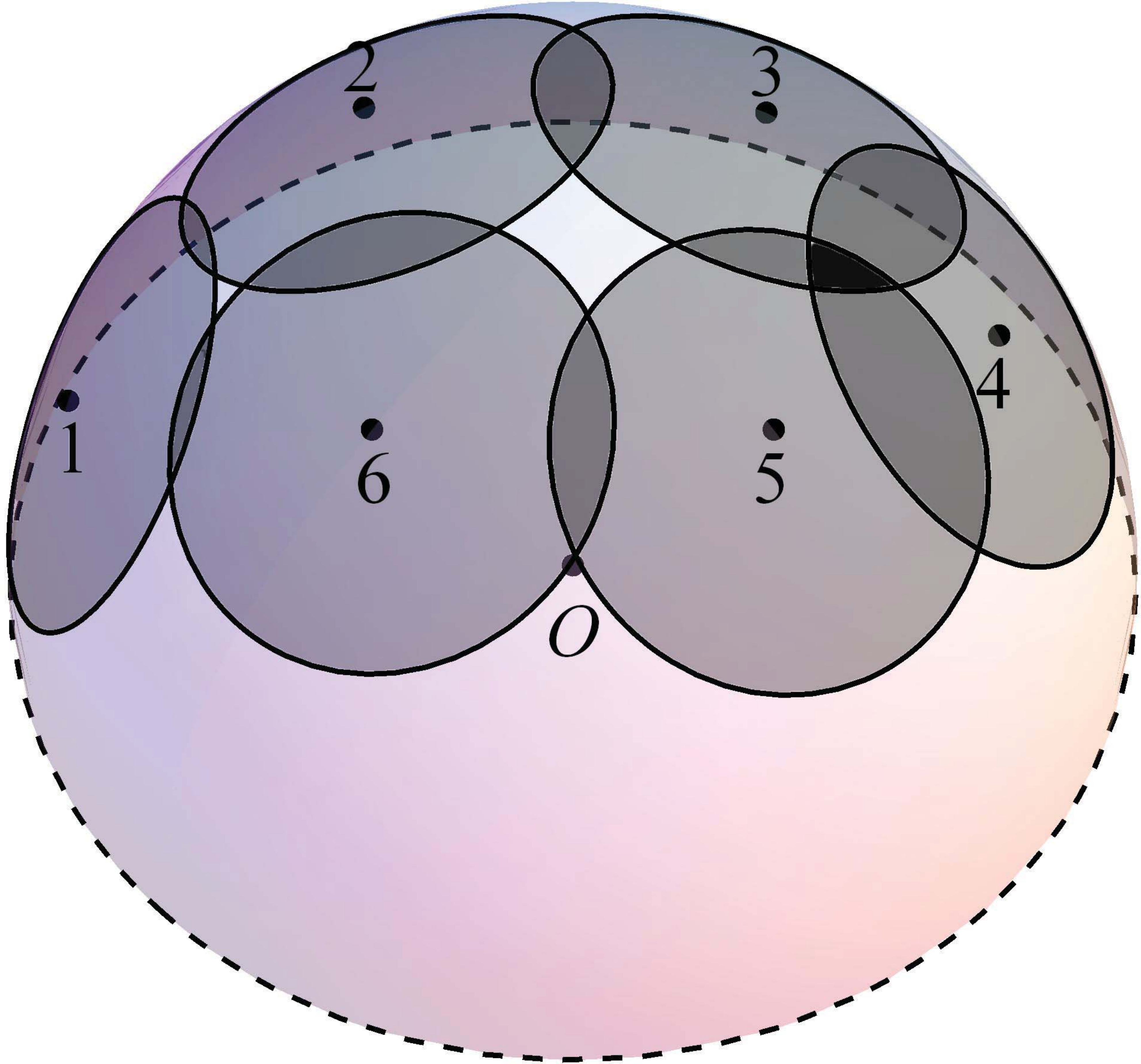}}
  \hspace{4pt}
  \subfloat[]{\includegraphics[width=0.2\textwidth]{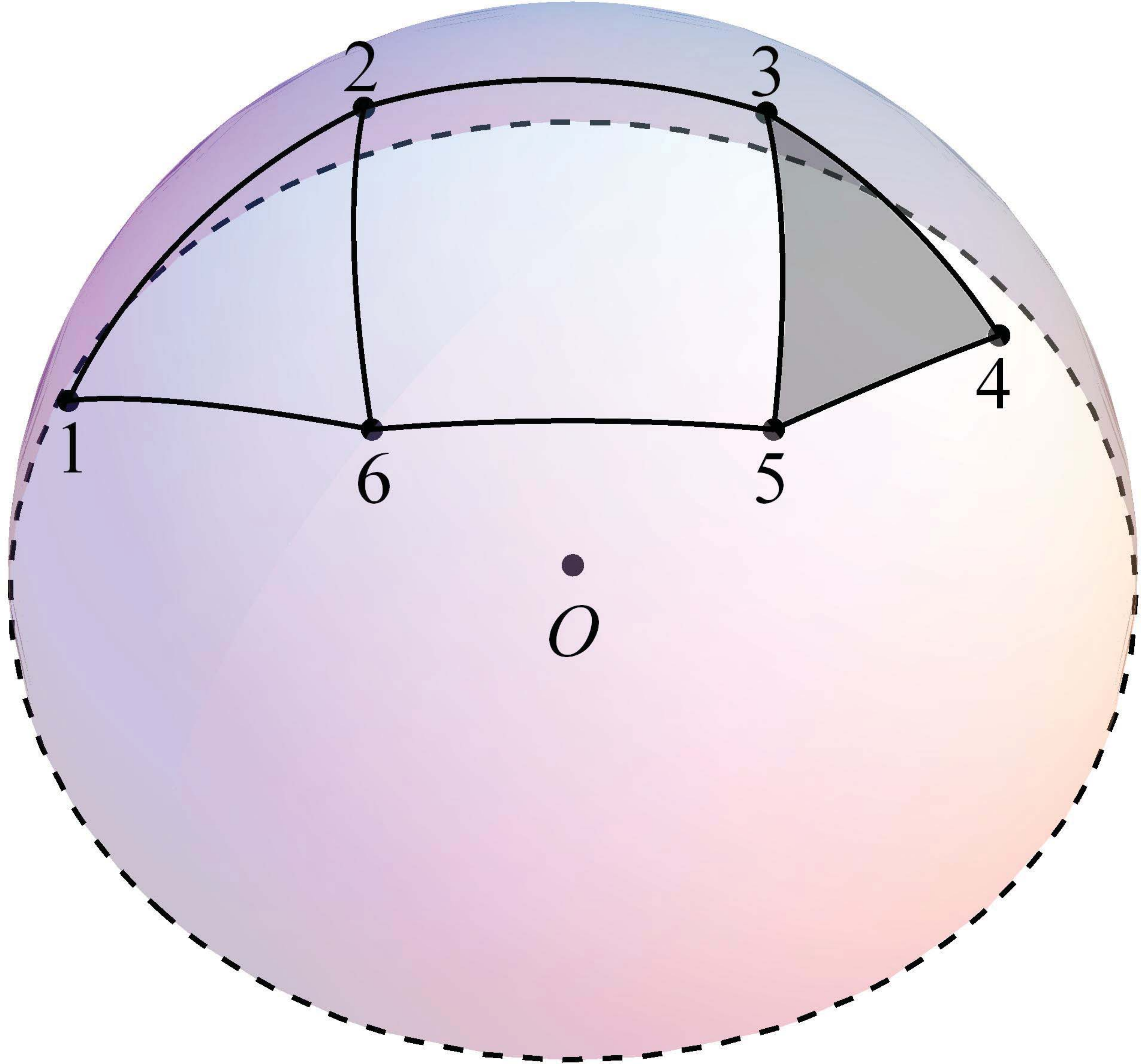}}
  \\
  \subfloat[]{\includegraphics[width=0.2\textwidth]{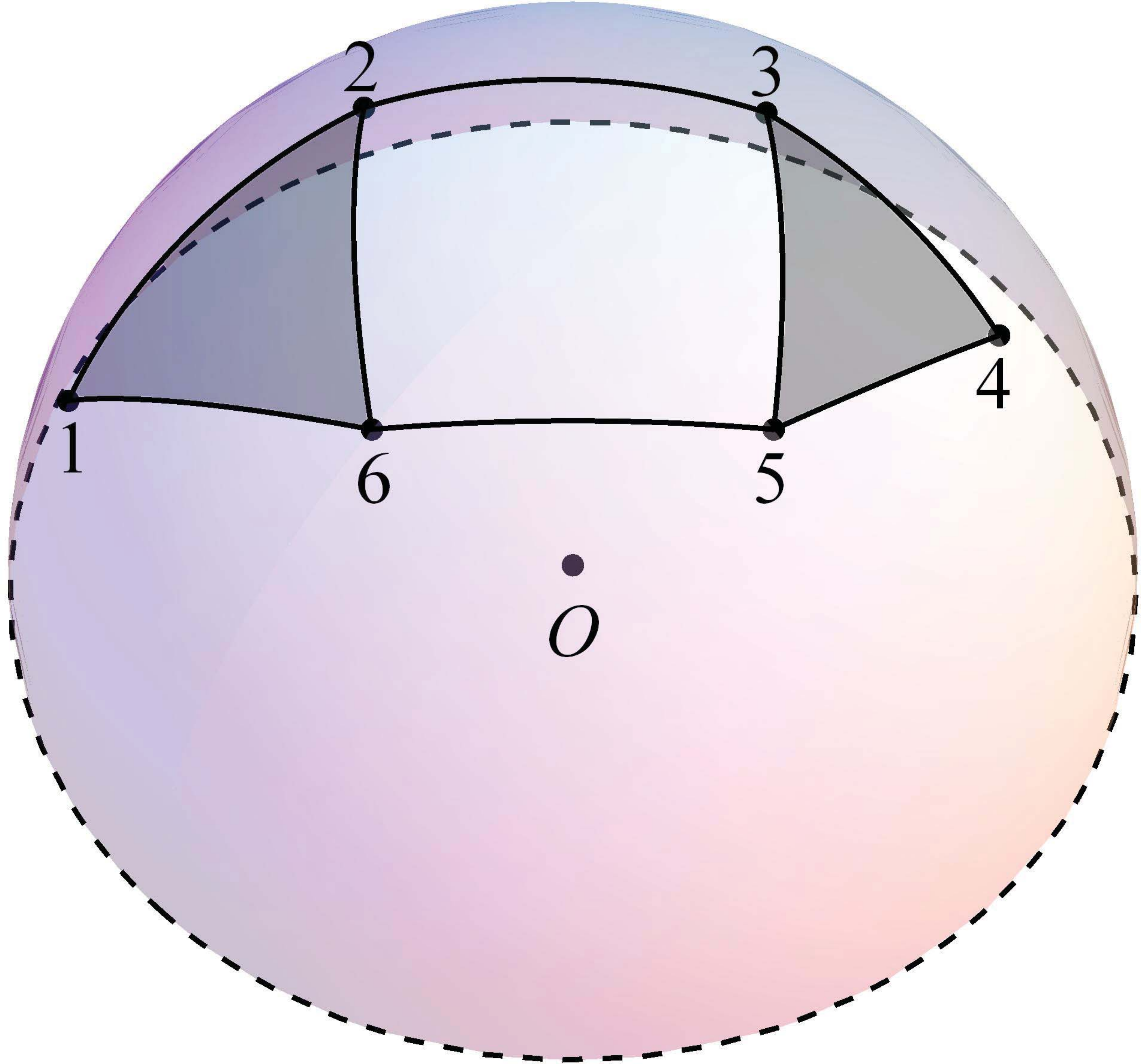}}
  \hspace{4pt}
  \subfloat[]{\includegraphics[width=0.2\textwidth]{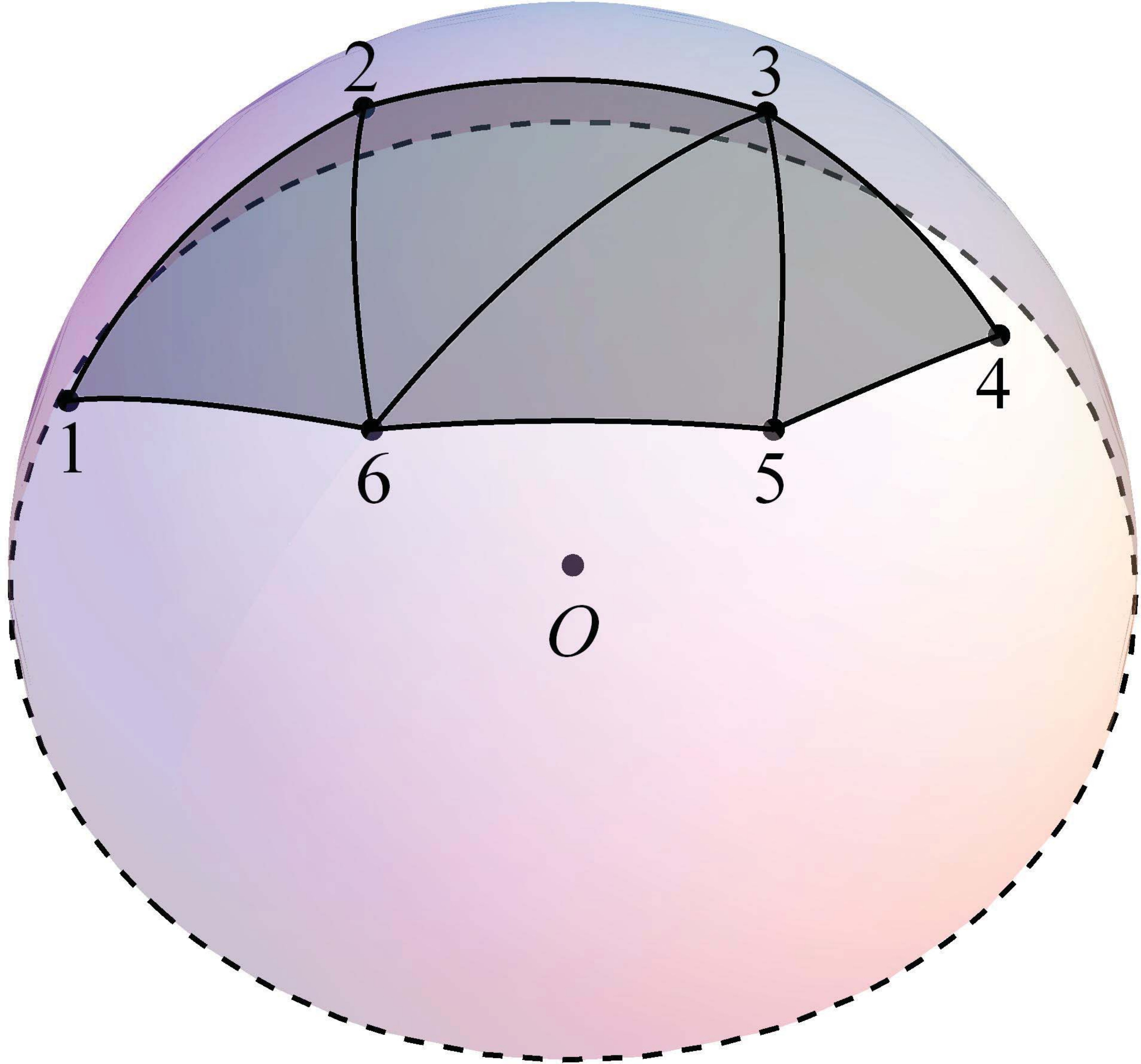}}
  \caption{(a) a WSN, (b) $\check{\textrm{C}}$ech complex, (c) Rips
    Complex under $R_c = 2 R_s$, (d) Rips Complex under $R_c = 2.5
    R_s$}
  \label{example}
\end{figure}

Figure~\ref{example} shows a WSN, its $\check{\textrm{C}}$ech complex and
two Rips complexes for two different values of $R_c$. Depending on the
relation of $R_c$ and $R_s$, the Rips complex and the $\check{\textrm{C}}$ech
complex may be close or rather different. In this example, for
$R_c=2R_s$, the Rips complex sees the hole surrounded by ${2,3,5,6}$
as in the $\check{\textrm{C}}$ech complex whereas it is missed in the
Rips complex for $R_c=2.5R_s$. At the same time, the true coverage
hole surrounded by ${1,2,6}$ is missed in both Rips complexes.

In fact, as proved in \cite{BT82}, any coverage hole can be found in 
$\check{\textrm{C}}$ech complex. Unfortunately, the construction of $\check{\textrm{C}}$ech complex 
is of very high complexity even if the precise location information of nodes is provided. So a more easily computable tool, Rips complex, is used. But Rips complex can not always capture all coverage holes. To be more specific, there exist following
relations between $\check{\textrm{C}}_{R_s}^{(2)}(\mathcal{V})$ and 
$\mathcal{R}_{R_c}^{(2)}(\mathcal{V})$.

\begin{lemma} \label{lemmaCR}
Let $\mathcal{V}$ denote the set of node locations in a WSN on 
$\mathbb{S}^2$ with radius $R$, all nodes have the same sensing
radius $R_s$ and communication radius $R_c$, $R_s \ll R, R_c \ll R$,
then
\begin{equation}\label{cech rips}
\begin{split}
 &\mathcal{R}_{R_c}^{(2)}(\mathcal{V}) \subset \check{\textrm{C}}_{R_s}^{(2)}(\mathcal{V}) \subset \mathcal{R}_{2R_s}^{(2)}(\mathcal{V}), \\
 &\emph{whenever} \hspace{4pt} R_c \leq R \arccos ([3\cos^2(R_s/R)-1]/2)
\end{split}
\end{equation} 
\end{lemma}

\begin{IEEEproof}
 See the Appendix \ref{app1}.
\end{IEEEproof}

According to (\ref{cech rips}), some relationships between $\check{\textrm{C}}$ech 
complex and Rips complex in terms of coverage hole can be derived as illustrated in 
the following corollaries.

\begin{corollary} \label{CRcase1}
When $R_c \leq R \arccos ([3\cos^2(R_s/R)-1]/2)$, if there is no hole in 
$\mathcal{R}_{R_c}^{(2)}(\mathcal{V})$, there must be no hole in 
$\check{\textrm{C}}_{R_s}^{(2)}(\mathcal{V})$.
\end{corollary}


\begin{corollary} \label{CRcase2}
When $R_c \geq 2R_s$, if there is a hole in $\mathcal{R}_{R_c}^{(2)}(\mathcal{V})$, 
there must be a hole in $\check{\textrm{C}}_{R_s}^{(2)}(\mathcal{V})$.
\end{corollary}


\begin{corollary} \label{CRcase3}
When $R \arccos ([3\cos^2(R_s/R)-1]/2) < R_c < 2R_s$, there is no guarantee
relation between $\check{\textrm{C}}_{R_s}^{(2)}(\mathcal{V})$ and $\mathcal{R}_{R_c}^{(2)}(\mathcal{V})$.
\end{corollary}

From Corollary \ref{CRcase1}, a sufficient condition for coverage verification
can be derived. From Corollary \ref{CRcase2}, we can find a necessary condition for 
the existence of a hole in $\check{\textrm{C}}_{R_s}^{(2)}(\mathcal{V})$. Corollary
\ref{CRcase3} indicates that when there is no hole in $\mathcal{R}_{R_c}^{(2)}(\mathcal{V})$,
it is possible that there is a hole in $\check{\textrm{C}}_{R_s}^{(2)}(\mathcal{V})$. When there
is a hole in $\mathcal{R}_{R_c}^{(2)}(\mathcal{V})$, it is also possible that $\check{\textrm{C}}_{R_s}^{(2)}(\mathcal{V})$ contains no hole. 
From these corollaries, it can be seen that when $R_c > R \arccos ([3\cos^2(R_s/R)-1]/2)$, 
$\mathcal{R}_{R_c}^{(2)}(\mathcal{V})$ may miss a hole in $\check{\textrm{C}}_{R_s}^{(2)}(\mathcal{V})$.
Furthermore, a hole in a $\check{\textrm{C}}_{R_s}^{(2)}(\mathcal{V})$
not seen in a $\mathcal{R}_{R_c}^{(2)}(\mathcal{V})$ must be bounded by a spherical triangle. 
Based on this observation, a formal definition of spherical triangular hole is given as follows.

\begin{definition}[Spherical triangular hole]\label{trihole}
For a pair of complexes $\check{\textrm{C}}_{R_s}^{(2)}(\mathcal{V})$ and 
$\mathcal{R}_{R_c}^{(2)}(\mathcal{V})$ of a WSN, a spherical triangular hole is an uncovered 
region bounded by a spherical triangle formed by three nodes $v_0, v_1, v_2$, where 
$v_0, v_1, v_2$ can form a 2-simplex which appears in $\mathcal{R}_{R_c}^{(2)}(\mathcal{V})$ 
but not in $\check{\textrm{C}}_{R_s}^{(2)}(\mathcal{V})$.
\end{definition}

According to Definition \ref{trihole}, it can be seen from Figure \ref{example}
that when $R_c = 2R_s$, there is one spherical triangular hole bounded by the spherical
triangle formed by nodes 1, 2 and 6. And when $R_c = 2.5R_s$, there are two additional
spherical triangular holes, bounded by spherical triangles formed by nodes
2, 3, 6 and 3, 5, 6 respectively.

A summary of the main notations is given in Table \ref{tabnot}.

\begin{table}[ht]
\caption{Main notations}
\label{tabnot}
\centering
\begin{tabular}{|c|p{6cm}|}
\hline
\bfseries symbols & \bfseries meaning\\
\hline
$R_s$ & sensing radius of each sensor\\ \hline
$R_c$ & communication radius of each sensor \\ \hline
$R$ & the radius of sphere where sensors are deployed\\\hline
$\mathcal{V}$ & the set of sensor locations \\ \hline
$s_v$ & the sensing range of the sensor located at $v$ \\ \hline
$\mathcal{S}$ & collection of sensing ranges of sensors in $\mathcal{V}$ \\ \hline
$\check{\textrm{C}}_{R_s}^{(2)}(\mathcal{V})$ & 2-dimensional  $\check{\textrm{C}}$ech complex of the WSN denoted by $\mathcal{V}$ \\ \hline
$\mathcal{R}_{R_c}^{(2)}(\mathcal{V})$ & 2-dimensional Rips complex of the WSN denoted by $\mathcal{V}$ \\ \hline
$\lambda$ & the intensity of Poisson point process \\ \hline
$p(\lambda)$ & the probability of any point on sphere being inside a spherical triangular hole \\\hline
$p_l(\lambda)$ & lower bound of $p(\lambda)$ \\ \hline
$p_u(\lambda)$ & upper bound of $p(\lambda)$ \\ \hline
$p_l^\prime(\lambda)$ & asymptotic lower bound of $p(\lambda)$ when $R \to \infty$\\ \hline
$p_u^\prime(\lambda)$ & asymptotic upper bound of $p(\lambda)$ when $R \to \infty$ \\ \hline
\end{tabular}
\end{table}

\section{Bounds on proportion of spherical triangular holes} \label{secbound}

In this section, the conditions under which any point on $\mathbb{S}^2$ with radius $R$
is inside a spherical triangular hole are first given. The proportion of the area of spherical
triangular holes is chosen as a metric for accuracy evaluation. Closed-form
expressions for lower and upper bounds of the proportion are derived. 
Finally, the asymptotic lower and upper bounds are investigated when the 
radius of sphere tends to infinity.

\subsection{Preliminary} \label{secboundpre}
\begin{lemma} \label{condition} For any point on $\mathbb{S}^2$, it
  is inside a spherical triangular hole if and only if the following two conditions are 
  satisfied:
  \begin{enumerate}
  \item the great circle distance between the point and its closest node is
    larger than $R_s$.
  \item the point is inside a spherical triangle: the convex hull of three nodes with
  pairwise great circle distance less than or equal to $R_c$.
  \end{enumerate}
\end{lemma}

\begin{lemma} \label{distance} If there exists a point $O$ which
  is inside a spherical triangular hole, then $R_s < R\arccos \sqrt{[1+2\cos(R_c/R)]/3} $. 
\end{lemma}

\begin{IEEEproof}
According to Definition \ref{trihole}, if there is a point $O$ inside a spherical 
triangular hole, then there exists a 2-simplex $\sigma \in \mathcal{R}_{R_c}^{(2)}(\mathcal{V})$
while $\sigma \notin \check{\textrm{C}}_{R_s}^{(2)}(\mathcal{V})$, so  
$\mathcal{R}_{R_c}^{(2)}(\mathcal{V}) \not \subset \check{\textrm{C}}_{R_s}^{(2)}(\mathcal{V})$.
According to (\ref{cech rips}), we have $R_c > R \arccos ([3\cos^2(R_s/R)-1]/2) \Rightarrow
R_s < R\arccos \sqrt{[1+2\cos(R_c/R)]/3}$.
\end{IEEEproof}

\begin{lemma} \label{closedist} Let $O$ be a point inside a spherical triangular hole
  and $l$ denote the great circle distance between $O$ and its closest neighbour, then
  $R_s < l \leq R\arccos \sqrt{[1+2\cos(R_c/R)]/3}$.
\end{lemma}

The proof is similar as that of Lemma \ref{lemmaCR}.

Since nodes are assumed to be distributed on $\mathbb{S}^2$ according to 
a homogeneous Poisson point process with intensity $\lambda$, any point
has the same probability to be inside a spherical triangular hole. 
This probability in a homogeneous setting is also equal to
the proportion of the area of spherical triangular holes. 

We use spherical coordinates $(R, \theta, \varphi)$ to denote points on $\mathbb{S}^2$ with radius $R$,
where $\theta$ is polar angle and $\varphi$ is azimuth angle. Without loss of generality, we
consider the probability of the point $N$ with spherical coordinates $(R, 0, 0)$ being inside a
spherical triangular hole. Since the communication radius of each sensor is 
at most $R_c$, only the nodes within $R_c$ from the point $N$ can
contribute to the spherical triangle which bounds a spherical triangular hole containing $N$. 
Therefore, we only need to consider the Poisson point
process constrained on the spherical cap $C(N, R_c)$ which is also a
homogeneous Poisson process with intensity $\lambda$, where $C(N, R_c)$
denotes the spherical cap centered at point $N$ and the maximum great circle
distance between $N$ and points on the spherical cap is $R_c$. We denote this
process as $\Phi$. In addition, $T(x, y, z)$ denotes the property that
the point $N$ is inside the spherical triangular hole
bounded by the spherical triangle with points $x, y, z$ as vertices. When $n_0, n_1,
n_2$ are points of the process $\Phi$, $T(n_0, n_1, n_2)$ is also used
to denote the event that the spherical triangle formed by the nodes $n_0,
n_1, n_2$ bounds a spherical triangular hole containing the point $N$. In addition,
we use $T'(n_0, n_1, n_2)$ to denote the event that the nodes
$n_0, n_1, n_2$ can not form a spherical triangle which bounds a spherical triangular hole 
containing the point $N$.

Let $\tau_0 = \tau_0(\Phi)$ be the node in the process $\Phi$ which is
closest to the point $N$. There are two cases for the point $N$ to be inside a
spherical triangular hole. The first case is that the node $\tau_0$ can contribute to
a spherical triangle which bounds a spherical triangular hole containing the point $N$. The
second case is that the node $\tau_0$ can not contribute to any spherical triangle
which bounds a spherical triangular hole containing the point $N$ but other three
nodes can form a spherical triangle which bounds a spherical triangular hole containing the 
point $N$. So the probability that the point $N$ is inside a spherical triangular hole 
can be defined as 
\begin{equation} \label{eqprob}
  \begin{split}
    p(\lambda) &= \mathrm{P}\{N \textrm{ is inside a spherical triangular hole}\} \\
    &= \mathrm{P}\{\bigcup_{\{n_0, n_1, n_2\} \subseteq \Phi} T(n_0, n_1, n_2)\} \\
    &= \mathrm{P}\{\bigcup_{\{n_1, n_2\} \subseteq \Phi
      \backslash \{\tau_0(\Phi)\}} T(\tau_0, n_1, n_2)\} + p_{sec}(\lambda) 
  \end{split}
\end{equation}
where 
\begin{displaymath}
    p_{sec}(\lambda) = \mathrm{P}\{\bigcup_{\{n_0,\cdots, n_4\} \atop \subseteq \Phi
            \backslash \{\tau_0(\Phi)\}} T(n_0, n_1, n_2)\mid T^\prime(\tau_0, n_3, n_4) \}
\end{displaymath}
\noindent denotes the probability of the second case. $p_{sec}(\lambda)$ is generally very
small and is obtained by simulations.

\subsection{Analytical lower and upper bounds} \label{secboundcase2}

As conjectured from Corollary \ref{CRcase1}, there exist spherical
triangular holes only in the case $R_c > R \arccos ([3\cos^2(R_s/R)-1]/2)$, 
so we only consider this case. The lower and upper bounds of $p(\lambda)$
are given as follows.

\begin{theorem} \label{trihole2}
	When $R_c > R \arccos([3\cos^2(R_s/R)-1]/2)$, 
	$p_l(\lambda) < p(\lambda) < p_u(\lambda)$,
	where 
	\begin{equation} \label{eqlower1}
	  \begin{split}
	  	    & p_l (\lambda) = 2\pi\lambda ^2 R^4\int_{R_s/R}^{\theta_{0u}} \sin \theta_0 d\theta_0 \int_{2\pi - \varphi_m(\theta_0)}^{2\varphi_m(\theta_0)} d\varphi_1 \int_{\theta_0}^{\theta_{1u}(\theta_0, \varphi_1)}  \\
	  	    & \sin \theta_1\times e^{-\lambda |C(N, R\theta_0)|} e^{-\lambda |S^+(\theta_0, \varphi_1)|}
	  	    (1 - e^{-\lambda |S^-(\theta_0, \theta_1, \varphi_1)|})d\theta_1
	  \end{split}
	\end{equation}
	and
	\begin{equation} \label{equpper1}
	  \begin{split}
	  	   &p_u (\lambda) =  2\pi\lambda ^2 R^4\int_{R_s/R}^{\theta_{0u}} \sin \theta_0 d\theta_0 \int_{2\pi - \varphi_m(\theta_0)}^{2\varphi_m(\theta_0)} d\varphi_1 \int_{\theta_0}^{\theta_{1u}(\theta_0, \varphi_1)} \\
	  	   & \sin \theta_1\times e^{-\lambda |C(N, R\theta_0)|} e^{-\lambda |S^+(\theta_0, \varphi_1)|}
	  	   	 (1 - e^{-\lambda |S^-(\theta_0, \theta_0, \varphi_1)|})d\theta_1 \\
	  	    & + p_{sec}(\lambda)
	  \end{split}
	\end{equation}
	and
	$\theta_{0u} = \arccos \sqrt{[1+2\cos(R_c/R)]/3}$
	  \begin{eqnarray} \label{eqphim}
		  \varphi_m(\theta_0) = 
		  \begin{cases}
			  \pi & \text{if $\frac{R_s}{R} < \theta_0 \leq \frac{R_c}{2R}$} \\
			  \arccos \frac{\cos\frac{R_c}{R} - \cos^2 \theta_0}{\sin^2 \theta_0} & \text{othewise}
		  \end{cases}
	  \end{eqnarray}
	  \begin{align}
	  	  &\theta_{1u}(\theta_0, \varphi_1) = \min \{\theta_{1u1}(\theta_0, \varphi_1), \theta_{1u2}(\theta_0, \varphi_1) \} \label{eqtheta1u} \\
	  	  &\theta_{1u1}(\theta_0, \varphi_1) = \arccos \frac{\cos(R_c/R)}{\sqrt{1-\sin^2\theta_0 \sin^2 \varphi_1}} \label{eqtheta1u1}\\ 
	  	  & \qquad \qquad \qquad + \arctan(\cos \varphi_1 \tan \theta_0) \nonumber \\
	  	  &\theta_{1u2}(\theta_0, \varphi_1) = \arccos \frac{\cos(R_c/R)}{\sqrt{1-\sin^2\theta_0 \sin^2 (\varphi_1-\varphi_m(\theta_0))}} \label{eqtheta1u2}\\
	  	  & \qquad \qquad \qquad + \arctan(\cos (\varphi_1-\varphi_m(\theta_0)) \tan \theta_0)\nonumber \\
	  	  &|C(N, R\theta_0)| = 2\pi R^2(1-\cos \theta_0) \label{eqs}\\
	  	  &|S^+(\theta_0, \varphi_1)| = \int_{2\pi - \varphi_m(\theta_0)}^{\varphi_1} \int_{\theta_0}^{\theta_{1u}(\theta_0, \varphi)} R^2 \sin \theta d\theta d\varphi \label{eqsplus}\\ 
	  	  &|S^-(\theta_0, \theta_1, \varphi_1)| = \int_{\varphi_{2l}}^{\varphi_m(\theta_0)} \int_{\theta_0}^{\theta_{2u}} R^2 \sin \theta_2 d\theta_2 d\varphi_2 \label{eqsminus}\\	  	  
	  	  &\varphi_{2l} = \varphi_1 - \arccos \frac{\cos (R_c/R) - \cos \theta_1 \cos \theta_0}{\sin \theta_1 \sin \theta_0} \nonumber \\	  
	  	     &\theta_{2u} = \min \{\theta_{1u1}, \theta_{2u2} \}\nonumber 
	  	 \end{align}
	  	 	  	  \begin{align}
	  	     &\theta_{2u2} = \arccos \Big [\cos(R_c/R)/\sqrt{1-\sin^2\theta_0 \sin^2 (\varphi_2 - \varphi_1)} \Big] \nonumber\\
	  	     & \qquad + \arctan(\cos (\varphi_2 - \varphi_1) \tan \theta_1)\nonumber
	  \end{align}
	  $p_{sec}(\lambda)$ is obtained by simulations\footnote{It is a non-trivial task to derive a closed-form expression for $p_{sec}(\lambda)$. Furthermore, we find that it is much less than the closed-form part in upper bound $p_u (\lambda)$ and it has little impact on the derived bound. We thus get it by simulations.}.
\end{theorem}

Since the proof is tedious, we only give the main steps of the proof.
Please refer to Appendix \ref{app2} for detailed computation. 
 
 For the lower bound, we only consider 
the first case that the closest node $\tau_0$ must contribute to a
spherical triangle which bounds a spherical triangular hole containing the point $N$.
The main idea is to first fix the closest node $\tau_0$, and then sequentially
decide the regions where the other two nodes may lie in, and finally
do a triple integral.

\begin{figure}[ht]
  \centering
  \includegraphics[width=0.4\textwidth]{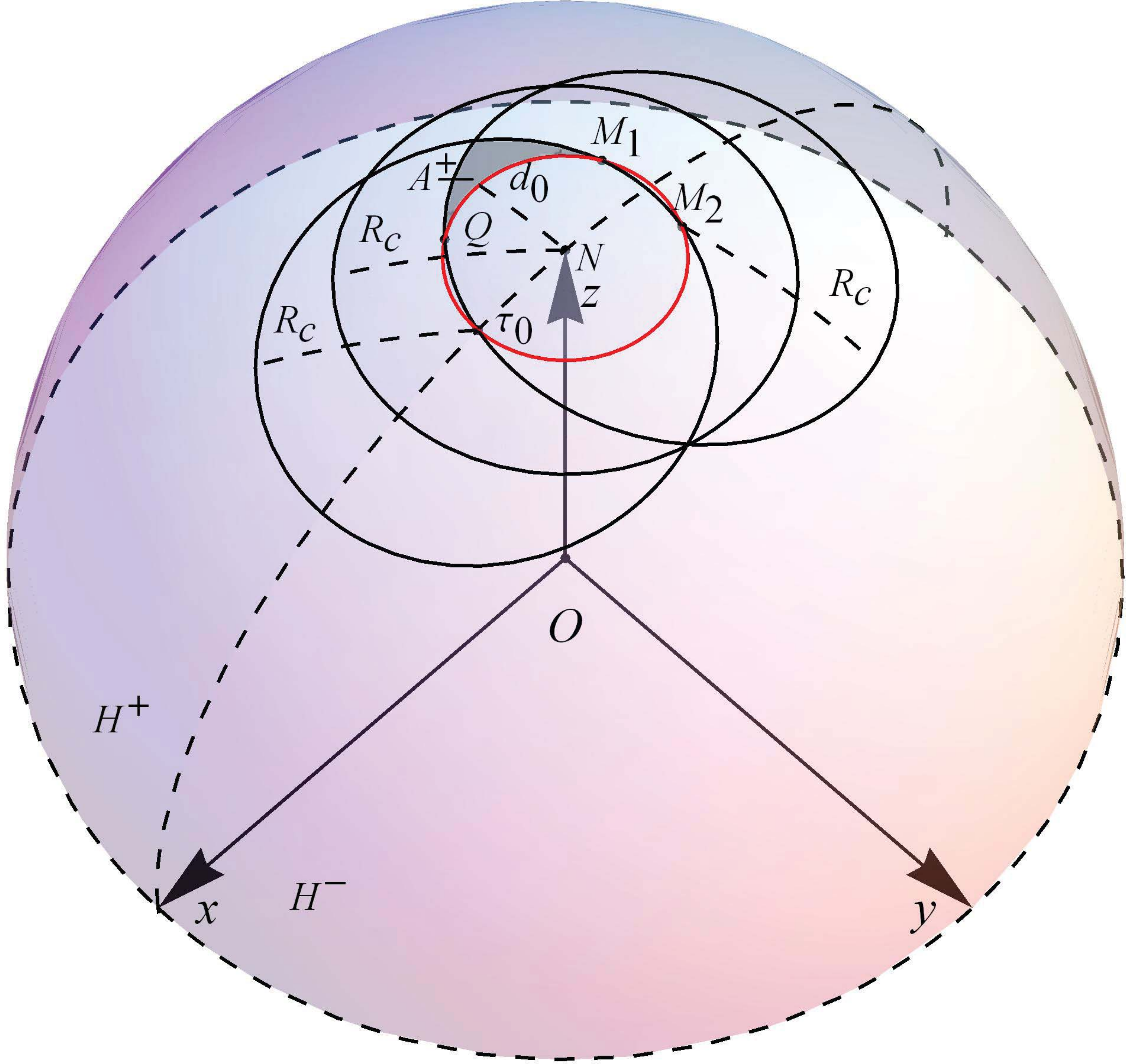}
  \caption{Illustration of region $A^+$ in the case $R \arccos([3\cos^2(R_s/R)-1]/2) < R_c \leq 2R_s$ 
  	or in the case $R_c > 2R_s, R_c/(2R) < \alpha_0 \leq \arccos \sqrt{[1+2\cos(R_c/R)]/3}$}
  \label{case2area1}
\end{figure}

\begin{figure}[ht]
  \centering
  \includegraphics[width=0.4\textwidth]{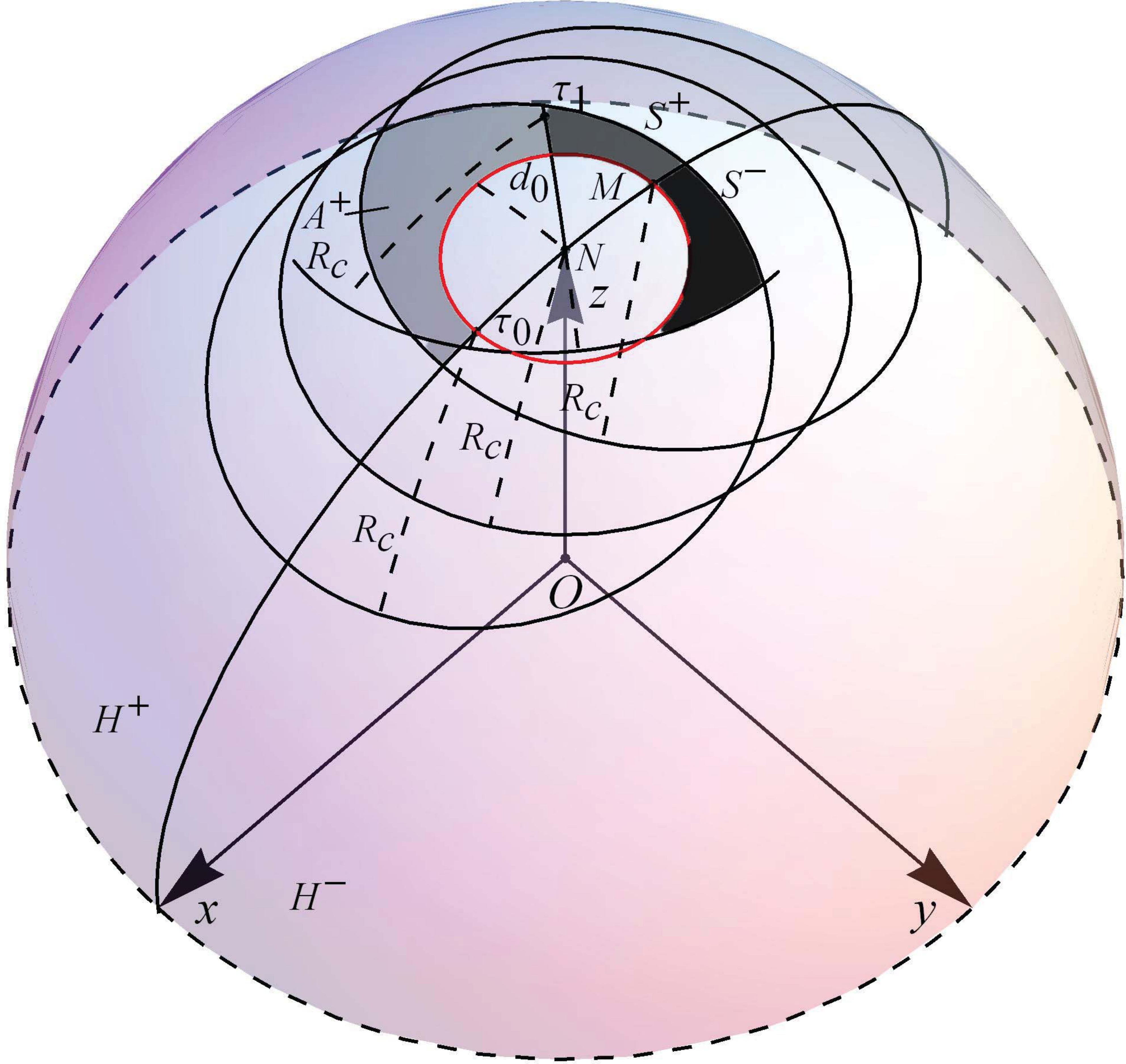}
  \caption{Illustration of regions $A^+, S^+$ and $S^-$ in the case $R_c > 2R_s, R_s/R < \alpha_0 \leq R_c/(2R)$ }
  \label{case3area1}
\end{figure}

Using spherical coordinates, we assume the closest node $\tau_0$ lies on 
$(R, \alpha_0, 0)$. Once the node $\tau_0$ is determined, the other two nodes must lie in 
the different half spaces: one in $H^+ = \mathbb{R}^+ \times (0, \pi/2) \times (\pi, 2\pi) $ 
and the other in $H^- = \mathbb{R}^+ \times (0, \pi/2) \times (0, \pi) $. 
Assume $n_1$ lies in $H^+$ and $n_2$ lies in $H^-$.
Since the great circle distance to $\tau_0$ is at most $R_c$,
$n_1$ and $n_2$ must also lie in the spherical cap $C(\tau_0, R_c)$. Furthermore, the
great circle distance to the point $N$ is at most $R_c$ and larger than $R\alpha_0$, they
should also lie in the region $C(N, R_c) \backslash C(N, R\alpha_0)$. Therefore, 
$n_1$ must lie in $H^+ \bigcap C(\tau_0, R_c) \bigcap C(N, R_c)
\backslash C(N, R\alpha_0) $ and $n_2$ must lie in $H^- \bigcap
C(\tau_0, R_c) \bigcap C(N, R_c) \backslash C(N, R\alpha_0)$. In addition,
considering the great circle distance between $n_1$ and $n_2$ should be at most
$R_c$ and the point $N$ should be inside the spherical triangle formed by $\tau_0$,
$n_1$ and $n_2$, $n_1$ must lie
in the shadow region $A^+$ shown in Figures \ref{case2area1} or \ref{case3area1}
under different situations.
In the case $R \arccos([3\cos^2(R_s/R)-1]/2) < R_c \leq 2R_s$
or in the case $R_c > 2R_s, R_c/(2R) < \alpha_0 \leq \arccos \sqrt{[1+2\cos(R_c/R)]/3}$, 
$A^+ = H^+ \bigcap C(\tau_0, R_c) \bigcap C(N, R_c)
\backslash C(N, R\alpha_0) \bigcap C(M_2, R_c)$, shown in Figure \ref{case2area1}. 
$M_1$ and $M_2$ are two intersection points between bases of spherical caps $C(N, R\alpha_0)$ and
$C(\tau_0, R_c)$. In the case $R_c > 2R_s, R_s/R < \alpha_0 \leq R_c/(2R)$, 
$A^+ = H^+ \bigcap C(\tau_0, R_c) \bigcap C(N, R_c)
\backslash C(N, R\alpha_0) \bigcap C(M, R_c)$, as in Figure \ref{case3area1}, where
$M$ is one intersection point between base of spherical caps $C(N, R\alpha_0)$ and 
the plane $xOz$.

\begin{figure}[ht]
  \centering
  \includegraphics[width=0.4\textwidth]{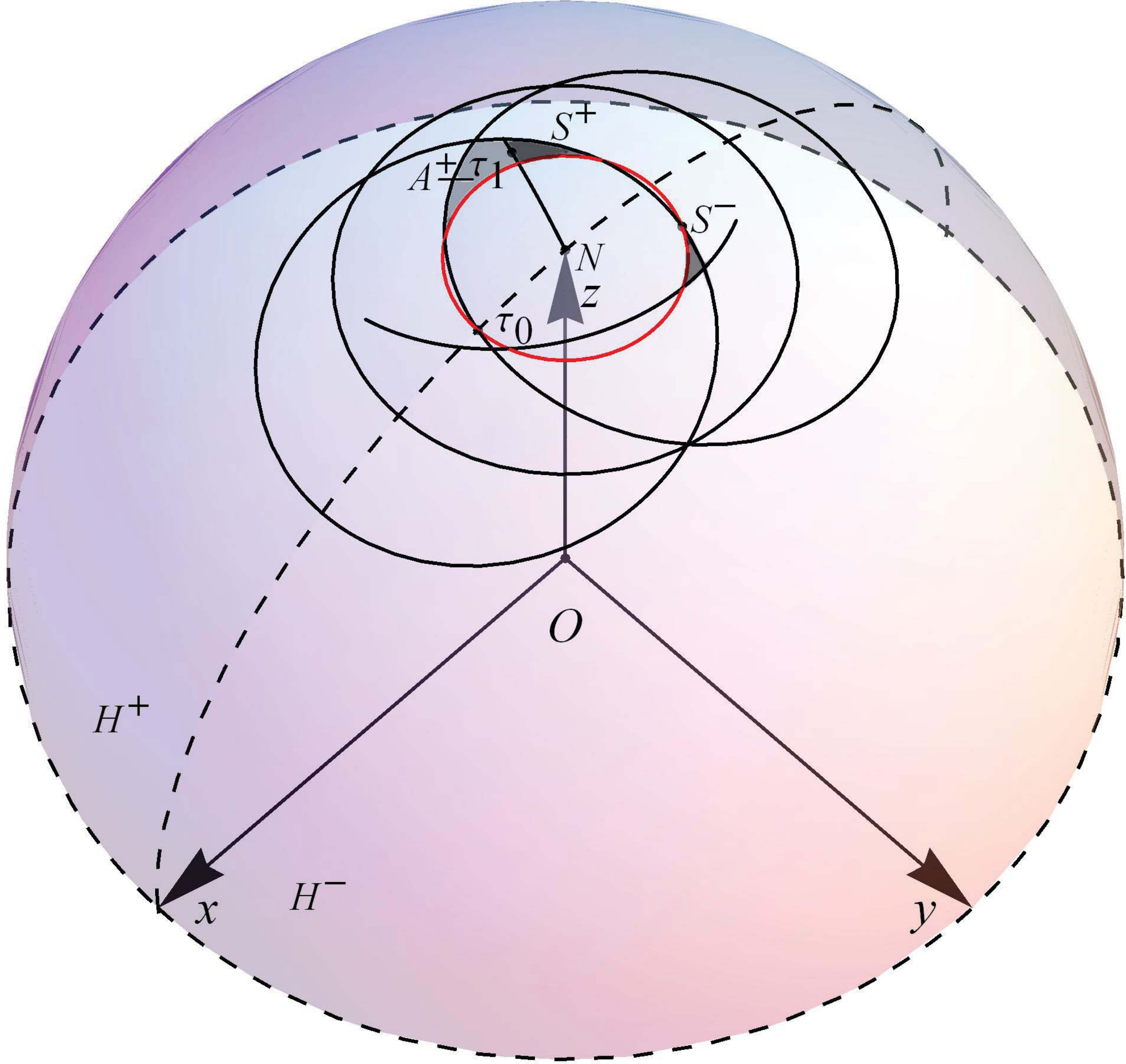}
  \caption{Illustration of regions $S^+$ and $S^-$ in the case $R \arccos([3\cos^2(R_s/R)-1]/2) < R_c \leq 2R_s$ }
  \label{case2area2}
\end{figure}

Ordering the nodes in $A^+$ by increasing azimuth angle so that $\tau_1 = (R, \theta_1, \varphi_1)$ 
has the smallest azimuth angle $\varphi_1$. And assume the nodes
$\tau_0$, $\tau_1$ and another node $\tau_2 \in H^- \bigcap C(\tau_0, R_c) \bigcap C(N, R_c)
\backslash C(N, R\alpha_0)$ can form a spherical triangle which bounds a spherical triangular hole containing
the point $N$, then $\tau_2$ must lie to the right of the great circle passing
through $\tau_1$ and $N$, denoted by $H^+(\varphi_1)$ which contains
all points with azimuth angle $\varphi \in (\varphi_1 - \pi, \varphi_1)$. In addition, the great circle
distance to $\tau_1$ is no larger than $R_c$, so the node $\tau_2$ must lie
in the region $S^-$, as illustrated in Figures ~\ref{case3area1} and \ref{case2area2}.

\begin{displaymath}
  \begin{split}
    & S^-(\tau_0, \tau_1)  =  S^- (\alpha_0, \theta_1, \varphi_1) =  H^- \bigcap C(\tau_0, R_c)  \\
    & \bigcap C(N, R_c) \backslash C(N, R\theta_0) \bigcap H^+(\varphi_1)
    \bigcap C(\tau_1, R_c)
  \end{split}
\end{displaymath}

\begin{figure}[ht]
  \centering \subfloat[]{\includegraphics[width=0.4\textwidth]{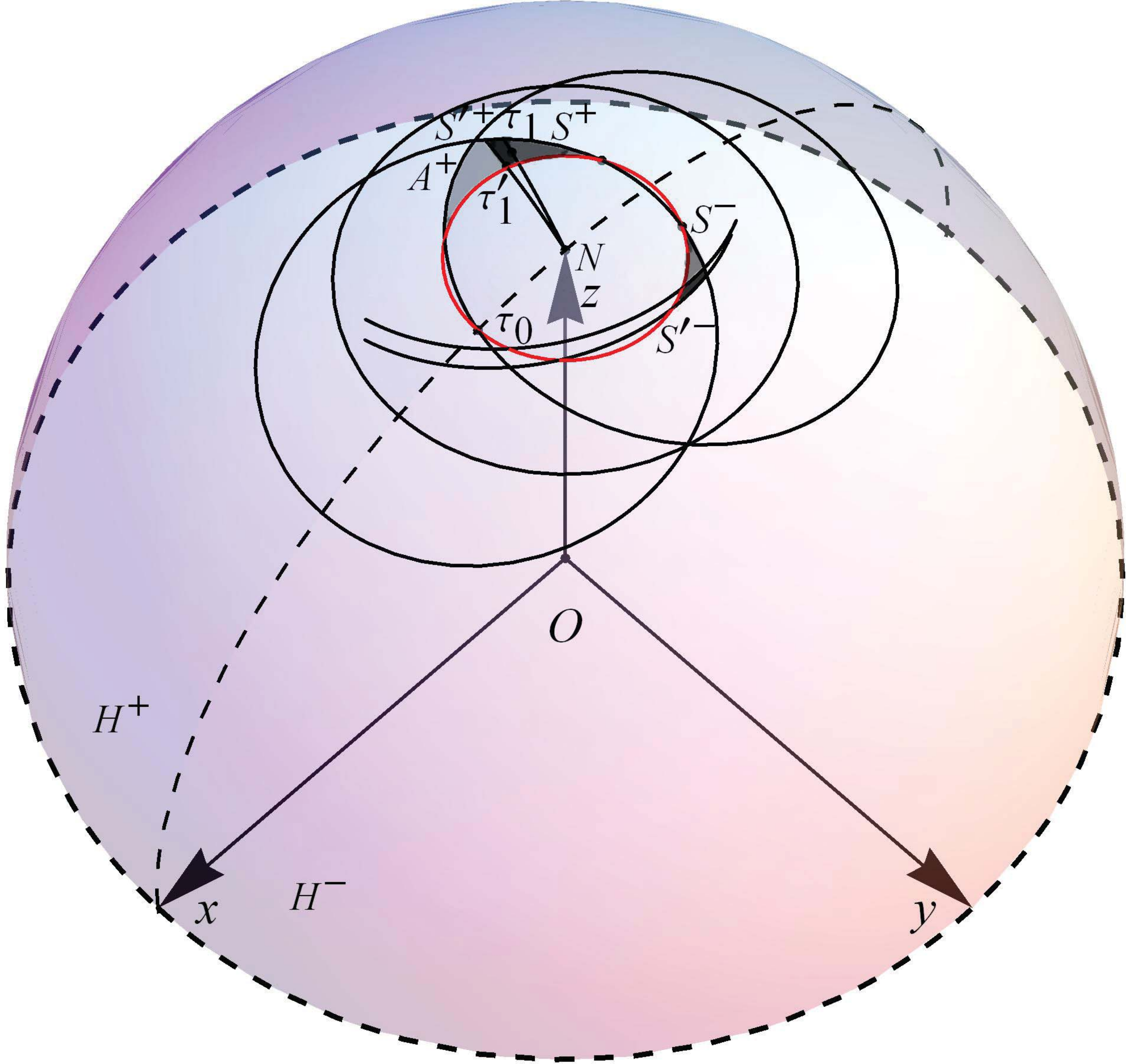}}
  \\
  \subfloat[]{\includegraphics[width=0.4\textwidth]{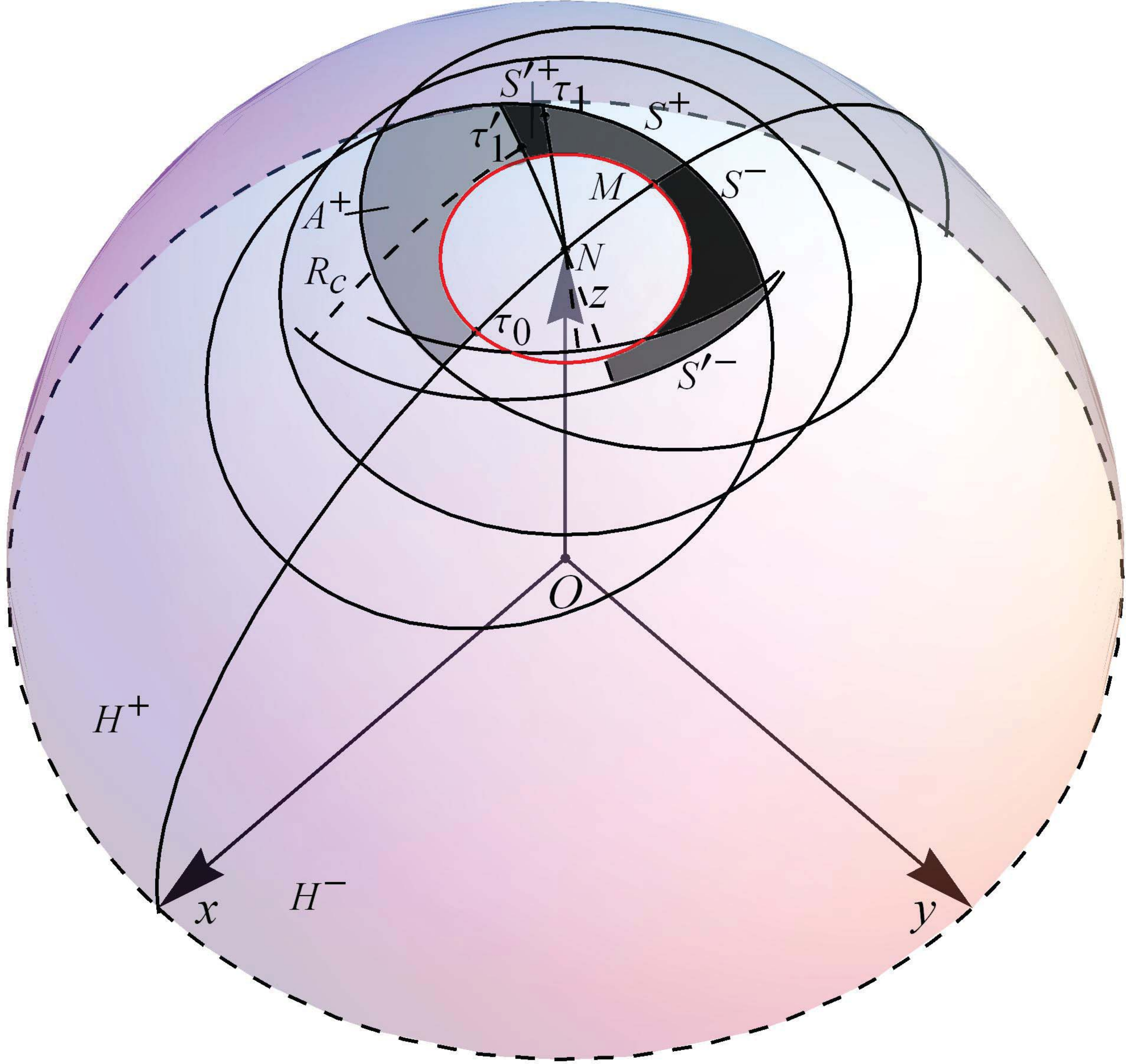}}
  \caption{Illustrations of regions $S^{'+}$ and $S^{'-}$ (a)
  	in the case $R \arccos([3\cos^2(R_s/R)-1]/2) < R_c \leq 2R_s$
  	or in the case $R_c > 2R_s, R_c/(2R) < \alpha_0 \leq \arccos \sqrt{[1+2\cos(R_c/R)]/3}$;
  	(b) in the case $R_c > 2R_s, R_s/R < \alpha_0 \leq R_c/(2R)$ }
  \label{case2area3}
\end{figure}

Assume only $\tau_0, \tau_1$ and nodes in $S^-(\tau_0, \tau_1)$ can
contribute to the spherical triangle which bounds a spherical triangular hole containing the
point $N$, we can get a lower bound of the probability that the
point $N$ is inside a spherical triangular hole. It is a lower bound because it is
possible that $\tau_1$ can not contribute to a spherical triangle which bounds a spherical  
triangular hole containing point $N$, but some other nodes with 
higher azimuth angles in the region $A^+$ can contribute to such a spherical triangle.
For example, in Figure \ref{case2area3}, if there is no node in $S^-$ but 
there are some nodes in $S^{\prime-}$, then $\tau_1$ can not contribute
to any spherical triangle which bounds a spherical triangular hole containing point $N$, but
$\tau_1^\prime$ can form such a spherical triangle with $\tau_0$ and another node in
$S^{\prime-}$.

%
%

Next we will prove the upper bound. As discussed in Section \ref{secboundpre},
there are two cases for the point $N$ being inside a spherical triangular hole. As for the 
second case that the closest node $\tau_0$ can not but some other nodes
can contribute to a spherical triangle which bounds a spherical triangular hole containing
the point $N$, it is not easy to obtain a closed-form expression for such
probability, so we get it by simulations. Simulation results show
that this probability is less than 0.16\% whenever $R_c \leq 3R_s$ with any 
intensity $\lambda$. So we still focus on the probability of the first
case.

Still consider the nodes in $A^+$, each node $(R, \theta, \varphi)$ corresponds to 
an area $|S^- (\alpha_0, \theta, \varphi)|$. The higher is the area $|S^- (\alpha_0, \theta, \varphi)|$, 
the higher is the probability that there is at least one node in 
$S^- (\alpha_0, \theta, \varphi)$, consequently the probability of the first case
will be higher. It can be seen from Figures \ref{case3area1} and \ref{case2area2} that the closer to 
$\alpha_0$ is $\theta$ and the closer to $\varphi_1$ is $\varphi$, the higher  
is the area  $|S^- (\alpha_0, \theta, \varphi)|$. So the largest area 
$|S^- (\alpha_0, \theta, \varphi)|$ is  $|S^- (\alpha_0, \alpha_0, \varphi_1)|$.
Based on that, the upper bound can be derived. 

As can be seen, the expression for lower bound is closed-form, while the expression for upper bound is not 
exactly closed-form since it includes a non-analytical part $p_{sec}(\lambda) $. As for lower bound and the 
closed-form part for upper bound, we use numerical integration to approximate the triple integrals. As for 
$p_{sec}(\lambda) $, we get it by simulations since it is very small, it has little impact on the derived bound.

\subsection{Asymptotic lower and upper bounds}

Intuitively, when $R \to \infty$, the case on sphere should be the same as 
that in plane, which is shown in the following theorem.

\begin{theorem} \label{theo2}
When $R \to \infty$ and $R_c > \sqrt{3}R_s$, lower and upper bounds in \emph{(\ref{eqlower1})} and \emph{(\ref{equpper1})}
become 
	\begin{equation} \label{eqlower2}
	  \begin{split}
	    p_l^\prime &(\lambda) =  2\pi\lambda ^2 \int_{R_s}^{R_c/\sqrt{3}} r_0 dr_0 \int_{\varphi_l(r_0)}^{\varphi_u(r_0)} d\varphi_1^\prime \int_{r_0}^{R_1(r_0, \varphi_1^\prime)} \\
	    	  	    & e^{-\lambda \pi r_0^2} \times e^{-\lambda |S^+(r_0, \varphi_1^\prime)|}
	    	  	    (1 - e^{-\lambda |S^-(r_0, r_1, \varphi_1^\prime)|})r_1 dr_1
	  \end{split}
	\end{equation}
	and 
	\begin{equation} \label{equpper2}
	  \begin{split}
	    p_u^\prime &(\lambda) =  2\pi\lambda ^2 \int_{R_s}^{R_c/\sqrt{3}} r_0 dr_0 \int_{\varphi_l(r_0)}^{\varphi_u(r_0)} d\varphi_1^\prime \int_{r_0}^{R_1(r_0, \varphi_1^\prime)} \\
	    	  	    & e^{-\lambda \pi r_0^2} \times e^{-\lambda |S^+(r_0, \varphi_1^\prime)|} (1 - e^{-\lambda |S^-(r_0, r_0, \varphi_1^\prime)|})r_1 dr_1\\
	    & + p_{sec}(\lambda)
	    \end{split}
	\end{equation}
	where 
	\begin{eqnarray} \label{eqphil}
	  \varphi_l(r_0) = 
	  \begin{cases}
		  0 & \text{if $R_s < r_0 \leq R_c/2$} \\
		  2\arccos(R_c /(2r_0)) & \text{othewise}
	  \end{cases}
	 \end{eqnarray}
	\begin{eqnarray} \label{eqphiu}
	 \varphi_u(r_0) = 
	 \begin{cases}
	  \pi & \text{if $R_s < r_0 \leq R_c/2$} \\
	  \pi - 4\arccos \frac{R_c}{2r_0} & \text{othewise}
	 \end{cases}
	 \end{eqnarray}
	\begin{align}
		& R_1(r_0, \varphi_1^\prime) = \min(\sqrt{R_c^2 - r_0^2\sin^2 \varphi_1^\prime} - r_0 \cos \varphi_1^\prime, \label{eqr1}\\
		& \qquad   \sqrt{R_c^2 - r_0^2\sin^2 (\varphi_1^\prime +\varphi_l(r_0))} + r_0 \cos (\varphi_1^\prime+\varphi_l(r_0))) \nonumber \\
		&|S^+(r_0, \varphi_1^\prime)| = \int_{\varphi_l(r_0)}^{\varphi_1^\prime} \int_{r_0}^{R_1(r_0, \varphi^\prime)} r dr d\varphi^\prime \label{eqsplusR}\\
		&|S^-(r_0, r_1, \varphi_1^\prime)| = \int_{\varphi_{2l}^\prime}^{-\varphi_l(r_0)} \int_{r_0}^{R_2(r_0, r_1, \varphi_1^\prime, \varphi_2^\prime)} r_2 dr_2 d\varphi_2^\prime \label{eqsminusR} \\
		&\varphi_{2l}^\prime = \varphi_1^\prime - \arccos(r_0^2+r_1^2-R_c^2)/(2r_0r_1) \nonumber\\
		&R_2(r_0, r_1, \varphi_1^\prime, \varphi_2^\prime) = \min(\sqrt{R_c^2 - r_0^2\sin^2 \varphi_2^\prime} - r_0 \cos \varphi_2^\prime, \nonumber  \\
		& \qquad \sqrt{R_c^2 - r_1^2\sin^2 (\varphi_2^\prime-\varphi_1^\prime)} + r_1 \cos (\varphi-\varphi_1^\prime))  \nonumber
	\end{align}
	$p_{sec}(\lambda)$ is obtained by simulations.
\end{theorem}

\begin{IEEEproof}
Please refer to Appendix \ref{app3}.
\end{IEEEproof}

Comparing (\ref{eqlower2}) and (\ref{equpper2}) to the results in the paper \cite{YMD12},
we can find that they are the same, which is quite logical since when $R \to \infty$ the local
of each node can be considered to be planar.

\section{Simulations and performance evaluation}

In this section, simulation settings are first given. Then simulation 
results are compared with analytical lower and upper bounds under 
different settings of $R_s, R_c, R$.

\subsection{Simulation settings}

A sphere centered at the origin with radius $R$ is considered in the simulations.
The probability of the point with spherical coordinate $(R, 0, 0)$ being inside a spherical triangular hole
is computed. Sensors are randomly distributed on the sphere 
according to a homogeneous Poisson point process with intensity $\lambda$. The sensing
radius $R_s$ of each node is set to be 10 meters and communication radius $R_c$
is chosen from 20 to 30 meters with interval of 2 meters. Let $\gamma = R_c/R_s$, then
$\gamma$ ranges from 2 to 3 with interval of 0.2. In addition, $\lambda$ is
selected from 0.001 to 0.020 with interval of 0.001. For each pair of $(\lambda, \gamma)$,
$10^7$ simulations are run to check whether the
point with spherical coordinate $(R, 0, 0)$ is inside a spherical triangular hole.

\subsection{Impact of $R_s$ and $R_c$}

\begin{figure}[ht]
  \centering \subfloat[]{\includegraphics[width=0.4\textwidth]{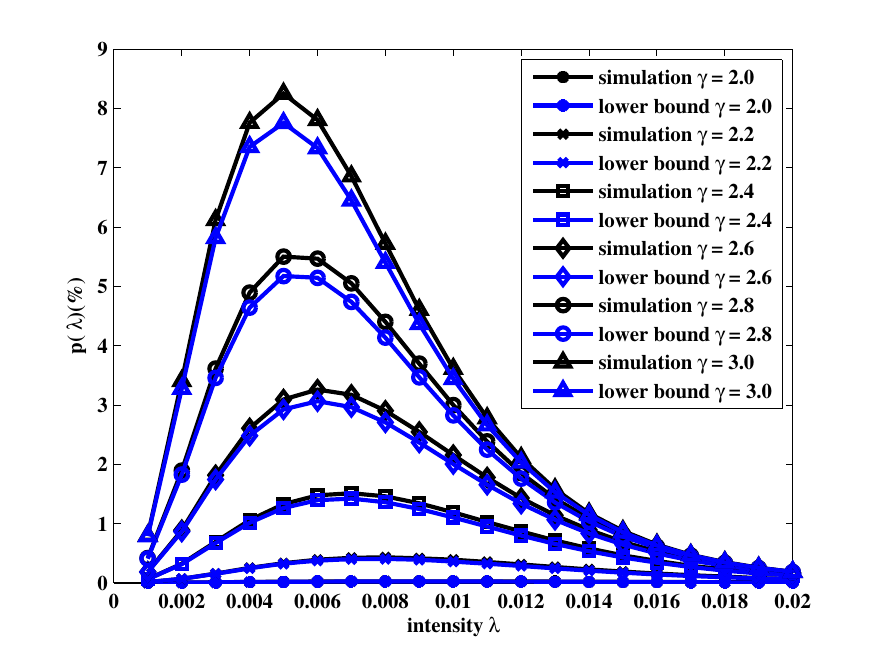}}
  \\
  \subfloat[]{\includegraphics[width=0.4\textwidth]{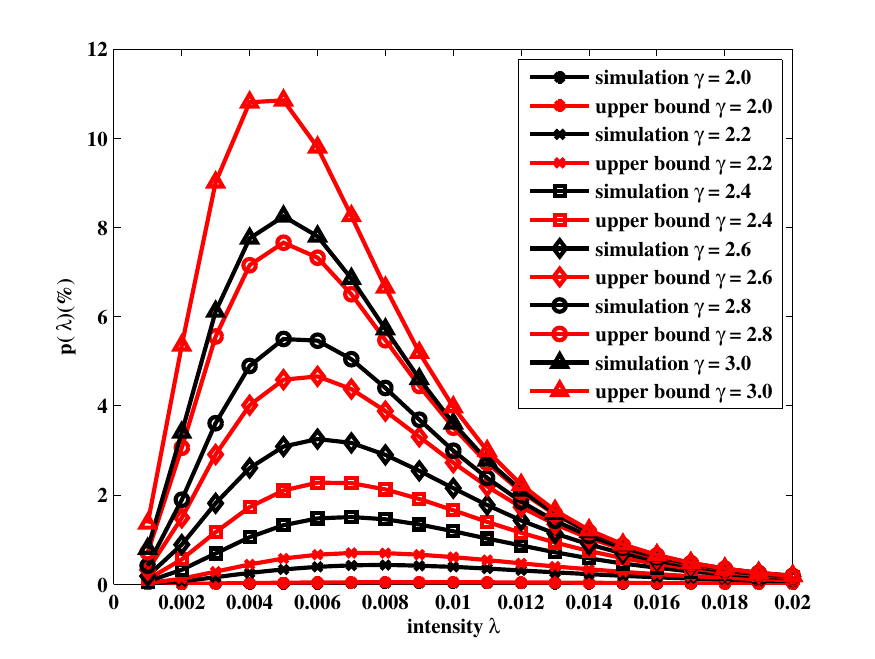}}\\
  \subfloat[]{\includegraphics[width=0.4\textwidth]{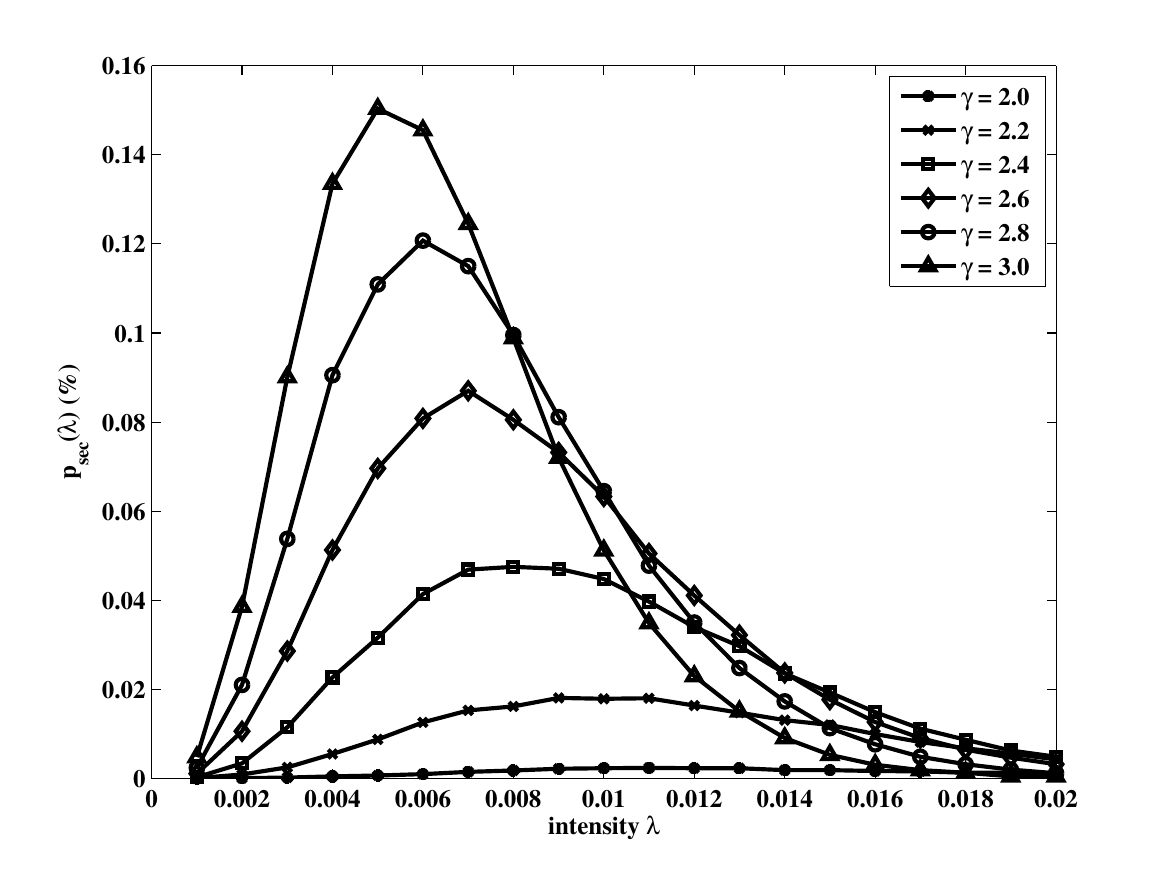}}
  \caption{Proportion of the area of spherical triangular holes under $R = 10 R_s$. (a) simulation results and
    lower bounds; (b) simulation results and upper bounds; (c) simulation results for $p_{sec}(\lambda)$}
  \label{figbound}
\end{figure}

As illustrated in Section \ref{secmod}, $R_s \ll R$ and $R_c \ll R$, here we fix
$R = 10 R_s$, choose $R_s$ to be 10 meters and $R_c$ to be 20 to 30 meters with interval of 2 meters, to analyse the impact of $R_s$ and $R_c$ on the probability of any point being inside a spherical
triangular hole. Under this configuration, the probability $p(\lambda)$ obtained by simulations is presented with the lower
and upper bounds in Figure \ref{figbound}(a) and \ref{figbound}(b) respectively. 
Note that the upper bounds contain the simulation results for $p_{sec}(\lambda)$
which are shown in Figure \ref{figbound}(c).

It can be seen that for any value of $\gamma$, $p(\lambda)$ has a maximum at a threshold value 
$\lambda_{c}$ of the intensity. As a matter of fact, for $\lambda \leq \lambda_{c}$, 
the number of nodes is small. Consequently the probability of any point being inside 
a spherical triangular hole is relatively small too. With the increase of $\lambda$, the 
connectivity between nodes becomes stronger. As a result, the probability of any point being 
inside a spherical triangular hole increases. However, when the intensity reaches 
the threshold value, the probability is up to its maximum.  $p(\lambda)$ 
decreases for $\lambda \geq \lambda_{c}$. The simulations also show  that $\lambda_{c}$ 
decreases with the increase of $\gamma$. 

On the other hand, it can be seen from Figure \ref{figbound}(a) and \ref{figbound}(b) that 
for a fixed intensity $\lambda$, $p(\lambda)$ increases with the increases of $\gamma$. 
That is because when $R_s$ is fixed, the larger $R_c$ is, the higher is the probability
of each spherical triangle containing a coverage hole. 

Furthermore, the maximum probability increases quickly with $\gamma$ ranging from 2.0 
to 3.0. These results can also provide some insights for planning of WSNs, which will
be discussed in Section \ref{discussion}.

Finally, it can be found in Figure \ref{figbound}(a) that the probability obtained by 
simulation is very well consistent with the lower bound. The maximum difference 
between them is about 0.5\%. Figure \ref{figbound}(b) shows that probability obtained 
by simulation is also consistent with the upper bound. The maximum difference 
between them is about 3\%.

\subsection{Impact of $R$}

\begin{figure}[!t]
  \centering \subfloat[]{\includegraphics[width=0.4\textwidth]{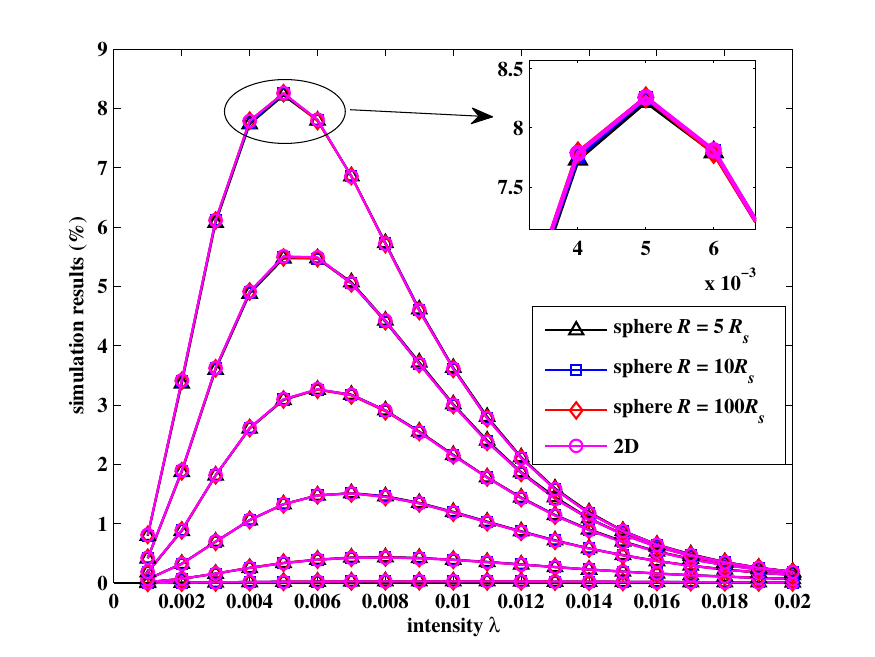}}
  \\
  \subfloat[]{\includegraphics[width=0.4\textwidth]{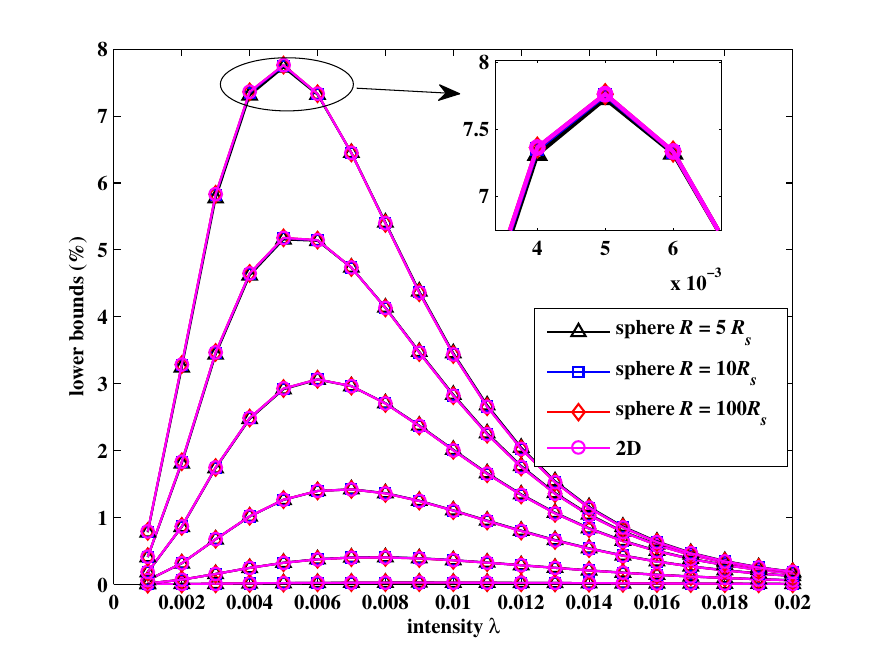}}
  \\
  \subfloat[]{\includegraphics[width=0.4\textwidth]{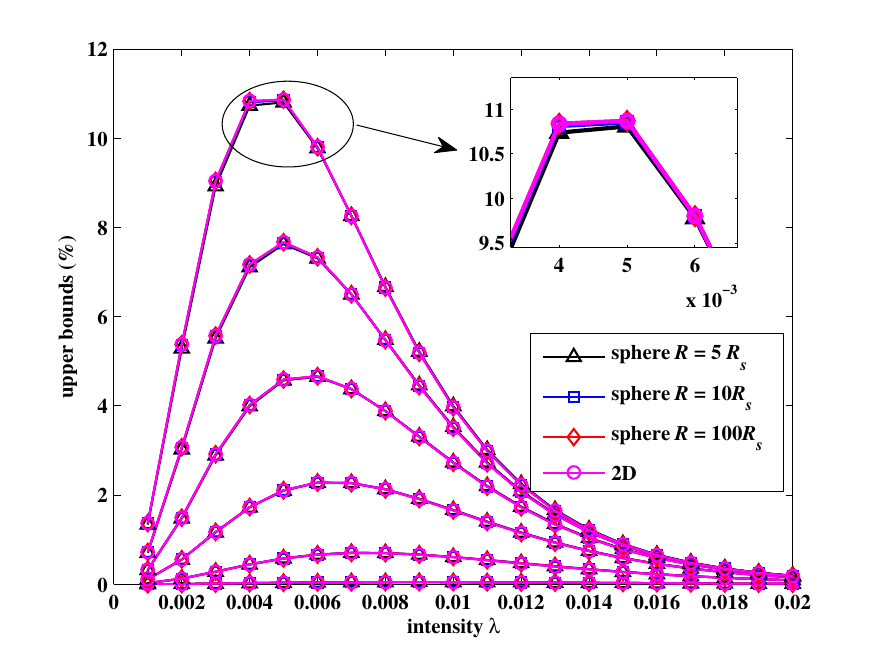}}
  \caption{Comparison of the proportion of the area of spherical triangular holes (a) comparison of simulation results;
    (b) comparison of lower bounds; (c) comparison of upper bounds}
  \label{figcomp}
\end{figure}

\begin{figure}[!ht]
  \centering
  \includegraphics[width=0.4\textwidth]{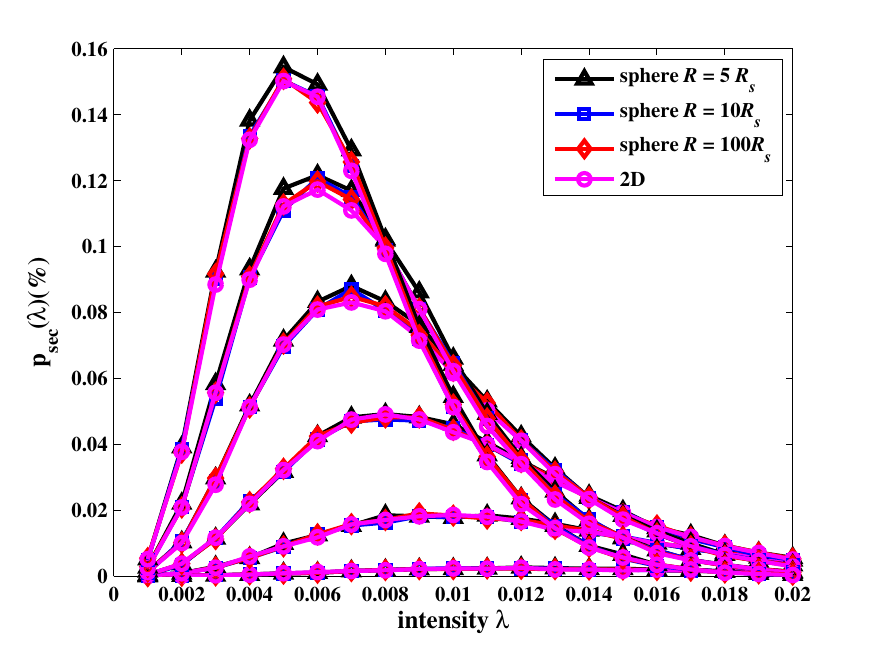}
  \caption{Simulation results for $p_{sec}(\lambda)$ }
  \label{figprobsec}
\end{figure}

Although we assume $R_s \ll R$ and $R_c \ll R$, to better understand the impact
of $R$ on the probability of any point being inside a spherical triangular hole,
we choose $R$ to be $5 R_s, 10 R_s$ and $100 R_s$. In these cases, $R_s$ is still 10 meters and $R_c$ is from 20 to 30 meters with interval of 2 meters. In addition, we also want to
know the difference of the probability under spherical and 2D planar cases.
Therefore, simulation results, lower and upper bounds of the probability
under spheres with radii $5 R_s, 10 R_s, 100 R_s$ and 2D plane are shown
in Figure \ref{figcomp}(a), \ref{figcomp}(b) and \ref{figcomp}(c) respectively.
Simulation results for $p_{sec}(\lambda)$ under spheres with radii 
$5 R_s, 10 R_s, 100 R_s$ and 2D plane are shown in Figure \ref{figprobsec}.
From Figure \ref{figprobsec}, we can find that $p_{sec}(\lambda)$ is less
than 0.16\% under any intensity in these cases.

It can be seen from Figure \ref{figcomp} that simulations results, lower and upper
bounds under spheres with radii $5 R_s, 10 R_s, 100 R_s$ and 2D plane are 
very close with each other. More precisely, the maximum difference of simulations results
under spheres with radii $5 R_s$ and $10 R_s$ is about 0.045\%, which is about 0.06\%
under spheres with radii $5 R_s$ and $100 R_s$ and is about 0.03\% under spheres
with radii $10 R_s$ and $100 R_s$. In addition, the maximum differences of simulation
results between 2D planar case and spherical cases with radii $5 R_s, 10 R_s, 100 R_s$
are 0.05\%, 0.03\% and 0.02\% respectively. It means the larger the radius of sphere is,
the more closer are the simulation results under sphere and 2D plane, it is because
the larger the radius of sphere is, the more likely of the local of each node on the sphere
to be planar.

With respect to lower and upper bounds, it is found that under any two 
spheres with radii $5 R_s, 10 R_s, 100 R_s$, the maximum difference of
lower and upper bounds are 0.06\% and 0.12\%  respectively. Furthermore,
under spheres with radii $5 R_s, 10 R_s, 100 R_s$ and 2D plane, 
the maximum difference of lower bounds is also 0.06\%, and that of upper bounds
is also 0.12\%. More importantly, under sphere with radius $100 R_s$ and 2D plane,
the maximum difference of lower bounds is $5 \times 10^{-6}$ and that of
upper bounds is $2.5 \times 10^{-5}$. It means the probabilities under cases of sphere with radius
$100 R_s$ and 2D plane are nearly the same, which is quite logical since when
the radius of sphere is much more larger than the sensing radius of any node,
the local of any node can be considered to be planar.

It can be further found that under above cases, the maximum differences of simulation results,
lower and upper bounds are all so small that they can be neglected. Consequently,
it also means that the radius of sphere has little impact on the probability of any 
point on the sphere to be inside a spherical triangular hole. 

\subsection{Discussions on applications} \label{discussion}

In this paper, we only consider spherical
triangular holes, for non-spherical triangular holes, we assume
they can be detected and covered by additional nodes. Under this
assumption, our analytical results can be used for planning 
of WSNs. For example, a WSN is used to monitor a mountain and 
the ratio $\gamma = 2$, according to the analytical upper bounds, 
we can see that the maximum proportion of the area of spherical triangular
holes under $\gamma = 2$ is about 0.06 \%, which can be neglected.
It means that as long as the surface of mountain can be 
spherically triangulated by nodes, we can say the mountain
is covered. But if $\gamma = 3$ and at least 95\% of the surface of 
the mountain should be covered, then it means that the proportion
of the area of spherical triangular holes can be at most 5\%.
From the analytical upper bounds of $\gamma = 3$, it can be 
seen that when the intensity $\lambda = 0.009$, the upper bound
is about 5\%, so in order to cover at least 95\% of the mountain,
the intensity of nodes should be larger than 0.009.

\section{Conclusions}

This paper studied the accuracy of homology-based coverage hole detection for wireless
sensor networks on sphere. First, the case when Rips complex may miss coverage
holes was identified. It was found that a hole missed by Rips complex must be 
bounded by a spherical triangle and a formal definition of spherical triangular
hole was given. Then we chose the proportion of the area of spherical triangular
holes as a metric to evaluate the accuracy. Closed-form expressions for
lower and upper bounds were derived. Asymptotic lower and upper bounds
are also investigated when the radius of sphere tends to infinity. 
Simulation results are well consistent with
the derived lower and upper bounds, with maximum differences of 0.5\% and 3\% respectively.
In addition, simulation results also show that the radius of sphere has little 
impact on the accuracy as long as it is much larger than communication and sensing 
radii of each sensor. This means that our results may be potentially applied to more general 3D surfaces
although the results are derived on sphere. This problem will be investigated in our future work.


%

\appendices
\section{Proof of Lemma \ref{lemmaCR}} \label{app1}

\begin{IEEEproof}
The second inclusion is trivial because for any $k$-simplex 
$[v_0,v_1,\cdots, v_k]\in \check{\textrm{C}}_{R_s}^{(2)}(\mathcal{V})$, 
it means the sensing ranges of these nodes have a common intersection,
so the pairwise distance $d(v_i,v_j) \leq 2R_s$ for all $0\leq i < j \leq k$,
which means  $[v_0,v_1,\cdots, v_k]\in \mathcal{R}_{2R_s}^{(2)}(\mathcal{V})$.

As for the first inclusion, it is clear that $\mathcal{R}_{R_c}^{(2)}(\mathcal{V})$
and $\check{\textrm{C}}_{R_s}^{(2)}(\mathcal{V})$ contain the same 0-simplices.
It is also easy to see that all 1-simplices
in $\mathcal{R}_{R_c}^{(2)}(\mathcal{V})$ must also be in 
$\check{\textrm{C}}_{R_s}^{(2)}(\mathcal{V})$ since for any 1-simplex $[v_i,v_j]$
with distance $d(v_i,v_j)\leq R_c \leq R \arccos ([3\cos^2(R_s/R)-1]/2) <
R \arccos(2\cos^2(R_s/R)-1) = 2R_s$, it means that the sensing ranges of the two nodes
have a common intersection. So we only need to prove that all 2-simplices in 
$\mathcal{R}_{R_c}^{(2)}(\mathcal{V})$ must be in $\check{\textrm{C}}_{R_s}^{(2)}(\mathcal{V})$.
It is equivalent to say that for any three nodes with pairwise great circle distance no larger than
$R_c$, their sensing ranges must have a common intersection. 

Assume a 2-simplex $[v_0,v_1,v_2] \in \mathcal{R}_{R_c}^{(2)}(\mathcal{V})$,
then the three nodes $v_0, v_1$ and $v_2$ must determine a plane $\alpha$. We consider
the spherical cap on $\mathbb{S}^2$ cut off by the plane $\alpha$. Since 
$R_c < R$, the spherical cap must be on a hemisphere. It is easy
to see that the intersection of the plane $\alpha$ and sphere $\mathbb{S}^2$
is a circle $c$. Let $O_1$ be the center of circle $c$, $O$ be the center of $\mathbb{S}^2$,
$P$ be the intersection of line $OO_1$ and $\mathbb{S}^2$.

Using spherical coordinates, we assume the point $P$ has a spherical coordinate $(R, 0, 0)$.
$P$ may be inside\footnote{It also includes the case that $P$ is on one arc of the spherical
triangle $v_0v_1v_2$.} or outside the spherical triangle $v_0v_1v_2$, which is shown in
Figure \ref{sphericaltri}(a) and \ref{sphericaltri}(b) respectively. 

\begin{figure}[ht]
  \centering \subfloat[]{\includegraphics[width=0.25\textwidth]{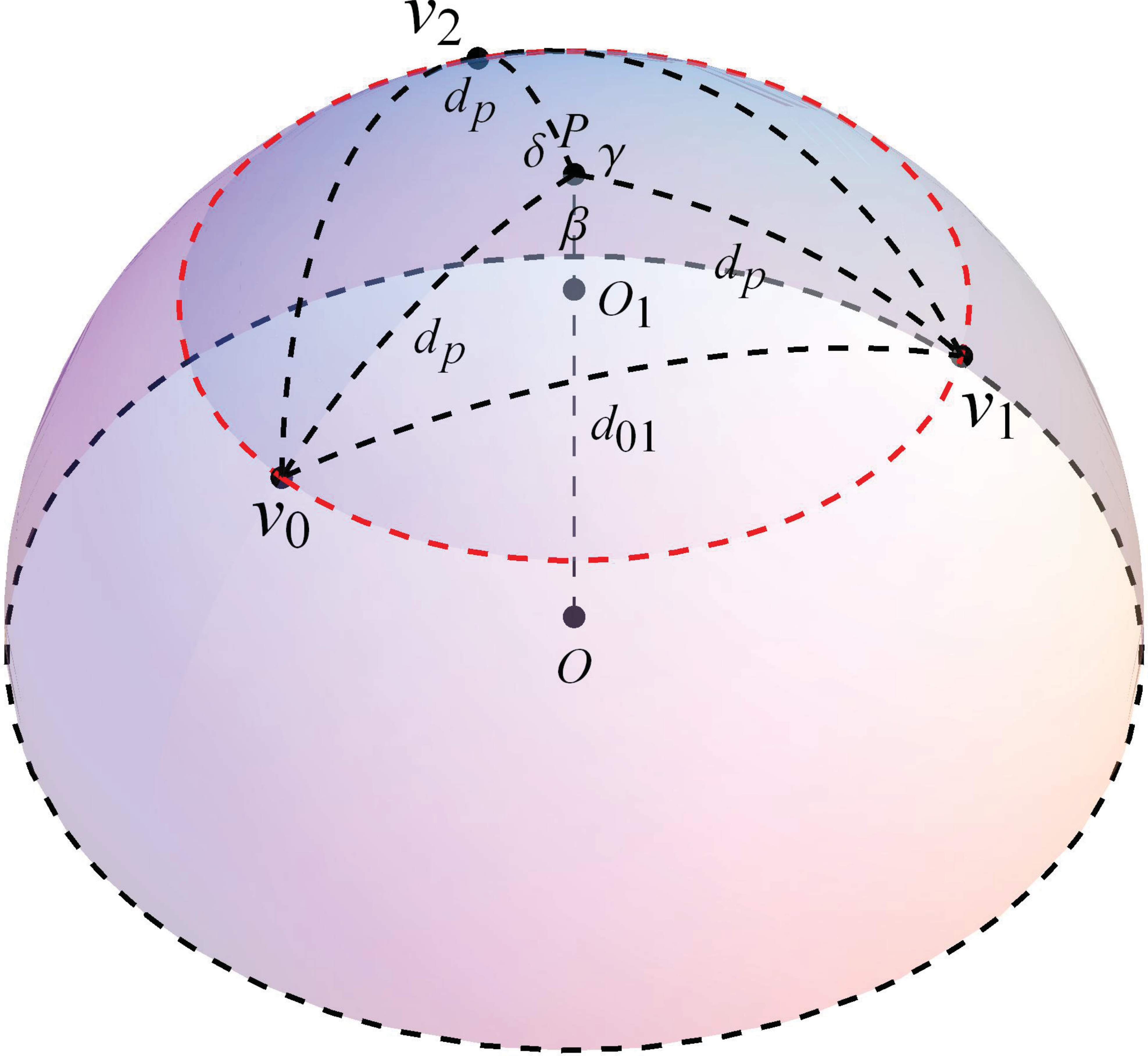}}  
  \subfloat[]{\includegraphics[width=0.25\textwidth]{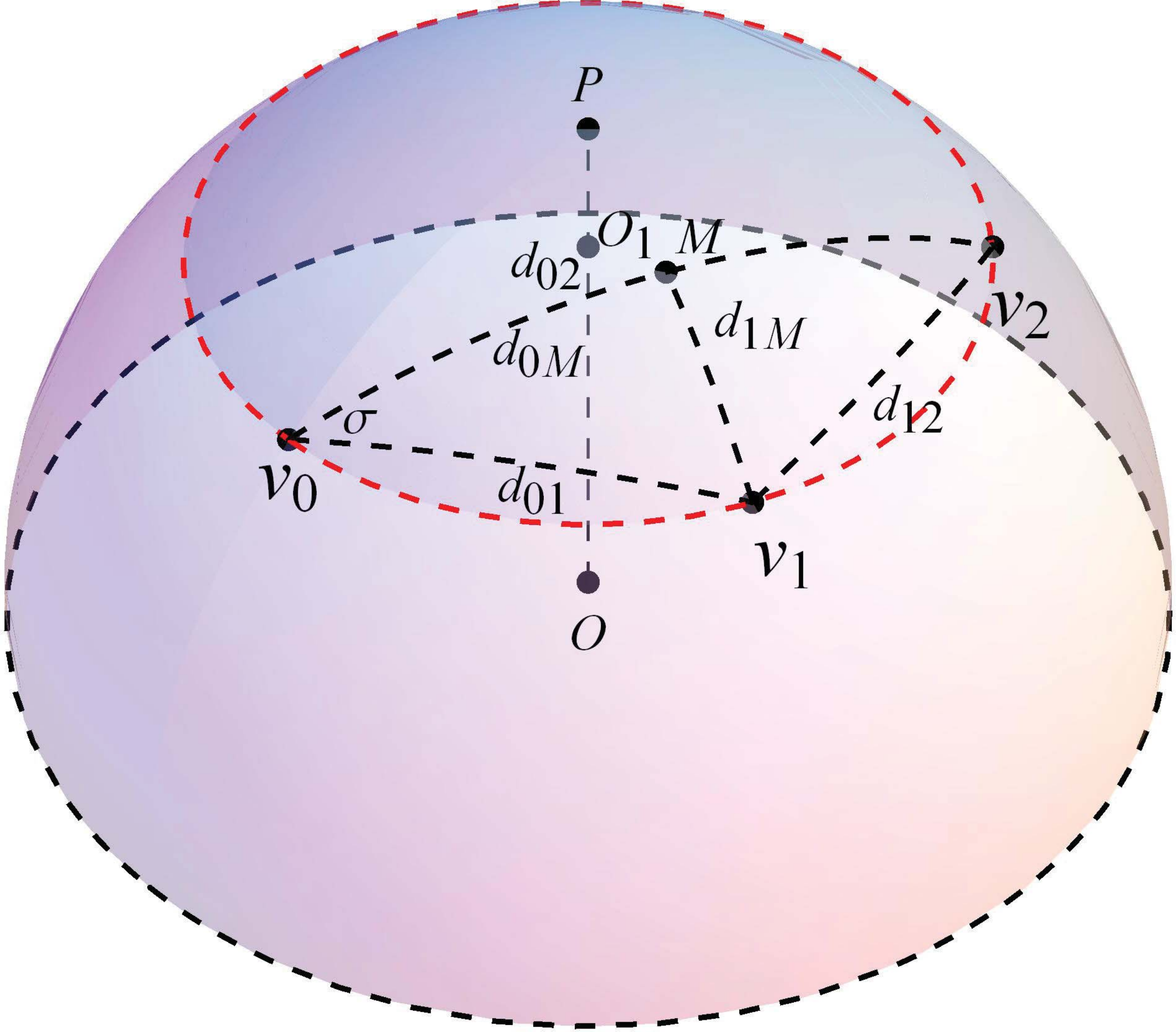}}
  \caption{Illustrations of $P$ and spherical triangle $v_0v_1v_2$: (a) $P$ is 
  inside the spherical triangle $v_0v_1v_2$; (b) $P$ is outside the spherical triangle $v_0v_1v_2$. }
  \label{sphericaltri}
\end{figure}

It can be seen that $P$ has the same 
great circle distance to $v_0, v_1$ and $v_2$, denoted by $d_p$. If $P$ is inside
the spherical triangle $v_0v_1v_2$, as shown in Figure \ref{sphericaltri}(a), then we can prove $d_p \leq R_s $. Since
$P$ lying inside the spherical triangle $v_0v_1v_2$ means $\beta + \gamma + \delta = 2\pi$,
there must be one angle no smaller than $2\pi/3$. Without loss of generality, assume
$\beta \geq 2\pi/3$. According to  the spherical law of consines, we have
$\cos(\beta) = \frac{\cos(d_{01}/R) - \cos^2(d_p/R)}{\sin^2(d_p/R)} \leq -1/2
\Rightarrow \cos (d_{01}/R) \leq [3\cos^2 (d_p/R) - 1]/2$. In addition,
$d_{01} \leq R_c \leq R \arccos ([3\cos^2(R_s/R)-1]/2) \Rightarrow 
\cos (d_{01}/R) \geq [3\cos^2(R_s/R)-1]/2$, and $0< d_{01}/R, d_p/R < \pi/2$, so we have
$[3\cos^2(R_s/R)-1]/2 \leq [3\cos^2 (d_p/R) - 1]/2 \Rightarrow 
d_p \leq R_s$, which means the point $P$ is a common intersection
of sensing ranges of $v_0, v_1$ and $v_2$, so $[v_0,v_1,v_2] \in \check{\textrm{C}}_{R_s}^{(2)}(\mathcal{V})$.

If $P$ is outside the spherical triangle $v_0v_1v_2$, as shown in Figure \ref{sphericaltri}(a), 
it indicates that the spherical triangle $v_0v_1v_2$
must be contained in half of the spherical cap. Assume $v_0, v_1$ and $v_2$ have spherical
coordinates $(R, \theta, \varphi_0), (R, \theta, \varphi_1)$ and $(R, \theta, \varphi_2)$, where
$\theta \in (0, \pi/2), \varphi_0 < \varphi_1 < \varphi_2$, then we have $\varphi_1 - \varphi_0, 
\varphi_2 - \varphi_1, \varphi_2 - \varphi_0 \in (0, \pi)$. Using $d_{01}, d_{12}, d_{02}$ to denote
the pairwise great circle distances between $v_0, v_1, v_2$, then according to 
the spherical law of consines, we have 
\begin{eqnarray}
\cos (d_{01}/R) = \cos^2 \theta + \sin^2 \theta \cos(\varphi_1 - \varphi_0) \label{eqapp1}\\
\cos (d_{12}/R) = \cos^2 \theta + \sin^2 \theta \cos(\varphi_2 - \varphi_1) \label{eqapp2}\\
\cos (d_{02}/R) = \cos^2 \theta + \sin^2 \theta \cos(\varphi_2 - \varphi_0) \label{eqapp3}
\end{eqnarray}

In addition, we use $\sigma$ to denote the angle between two arcs $\wideparen{v_0v_1}$ and $\wideparen{v_0v_2}$,
$M$ to denote the middle point of the arc $\wideparen{v_0v_2}$ and $d_{0M}, d_{1M}$ to 
denote great circle distances between $v_0$, $v_1$ and $M$. It can be seen $d_{0M} = d_{02}/2$.
Similarly, we have 
\begin{eqnarray}
 \cos \sigma = \frac{ \cos (d_{12}/R) - \cos (d_{01}/R) \cos (d_{02}/R)}{\sin (d_{01}/R) \sin (d_{02}/R)} \label{eqapp4} \\
 \cos \frac{d_{1M}}{R} = \cos \frac {d_{01}}{R} \cos \frac{d_{0M}}{2R} + \sin \frac{d_{01}}{R} \cos \frac{d_{0M}}{2R} \cos \sigma \label{eqapp5}
\end{eqnarray}

From (\ref{eqapp4}) and (\ref{eqapp5}), we can obtain 
\begin{equation}
 \cos \frac{d_{1M}}{R} = \frac{\cos (d_{01}/R) + \cos (d_{12}/R)}{2\cos (d_{02}/(2R))}
\end{equation}

Consequently 
\begin{equation} \label{eqapp6}
   \cos \frac{d_{1M}}{R} - \cos \frac{d_{0M}}{R} = \frac{\cos \frac{d_{01}}{R} + \cos \frac{d_{12}}{R} - \cos \frac{d_{02}}{R} - 1}{2\cos (d_{02}/(2R))} 
\end{equation}

From (\ref{eqapp1}), (\ref{eqapp2}), (\ref{eqapp3}) and (\ref{eqapp6}), we get
\begin{equation} \label{eqapp7}
 \cos \frac{d_{1M}}{R} - \cos \frac{d_{0M}}{R} = \frac{\sin^2 \theta \cos \frac{\varphi_2 - \varphi_0}{2} 
                                          \sin \frac{\varphi_1 - \varphi_0}{2} \sin \frac{\varphi_2 - \varphi_1}{2}}{\cos \frac{d_{02}}{2R}}
\end{equation}

Since $0< \varphi_1 - \varphi_0, \varphi_2 - \varphi_1, \varphi_2 - \varphi_0 < \pi$
and $0 < d_{1M}/R, d_{0M}/R , d_{02}/R < \pi/2$, it can be obtained from (\ref{eqapp7}) 
$d_{1M} < d_{0M} \leq R_c/2 < R_s$, which means the point $M$ is a common
intersection of the sensing ranges of $v_0, v_1$ and $v_2$, so 
$[v_0, v_1, v_2] \in \check{\textrm{C}}_{R_s}^{(2)}(\mathcal{V})$.
It means all 2-simplices in $\mathcal{R}_{R_c}^{(2)}(\mathcal{V})$ must be in $\check{\textrm{C}}_{R_s}^{(2)}(\mathcal{V})$.
Consequently the first inclusion is proved.

\end{IEEEproof}

\section{Proof of Theorem \ref{trihole2}} \label{app2}

\begin{IEEEproof}
We first prove the lower bound. It can be obtained from (\ref{eqprob}) that
\begin{displaymath} 
  \begin{split}
    p(\lambda) > \mathrm{P}\{\bigcup_{\{n_1, n_2\} \subseteq \Phi
      \backslash \{\tau_0(\Phi)\}} T(\tau_0, n_1, n_2)\} 
  \end{split}
\end{displaymath}

So for the lower bound, we only consider the first case that the closest node $\tau_0$ must contribute to a
spherical triangle which bounds a spherical triangular hole containing the point $N$.

Using spherical coordinates, we assume the closest node $\tau_0$ lies on 
$(R, \alpha_0, 0)$ and use $|S|$ to denote the area of the set $S$, then we can get the distribution of $\alpha_0$ as
\begin{equation} \label{eqlaw1}
   F_{\alpha_0}(\theta_0) = P(\alpha_0 \leq \theta_0) = 1 - e^{-\lambda|C(N, R\theta_0)|}
\end{equation}

\noindent since the event $\alpha_0 > \theta_0$ means that the spherical cap $C(N, R\theta_0)$
does not contain any nodes from the process, which is given by the void probability
$e^{-\lambda|C(N, R\theta_0)|}$. Furthermore, $|C(N, R\theta_0)|$ can be given as 
\begin{equation} \label{eqarea0}
 |C(N, R\theta_0)| =\int_0^{\theta_0} \int_0^{2\pi} R^2 \sin \theta d\varphi d\theta  = 2\pi R^2(1-\cos \theta_0)
\end{equation}

From (\ref{eqlaw1}) and (\ref{eqarea0}), we can get the density of $\tau_0$
\begin{equation} \label{eqdens0}
  F_{\alpha_0}(d\theta_0) = 2\pi \lambda R^2 \sin \theta_0 e^{-\lambda|C(N, R\theta_0)|}d\theta_0 
\end{equation}

The integration range for $\theta_0$ can be easily obtained. According to 
Lemma \ref{closedist}, we have $R_s < R\theta_0 \leq R \arccos \sqrt{[1+2\cos(R_c/R)]/3}$,
so $R_s/R < \theta_0 \leq \theta_{0u} = \arccos \sqrt{[1+2\cos(R_c/R)]/3}$.

Therefore the probability of the first case can be given as 
\begin{equation} \label{eqplower1}
  \begin{split}
    & \mathrm{P}\{\bigcup_{\{n_1, n_2\} \subseteq \Phi \backslash \{\tau_0(\Phi)\}} T(\tau_0, n_1, n_2)\} \\
    = & \int_{R_s/R}^{\theta_{0u}} \mathrm{P}\{\bigcup_{\{n_1, n_2\} \subseteq
      \Phi_{\theta_0}^\prime} T((R, \theta_0, 0), n_1, n_2)\} F_{\alpha_0}(d\theta_0)
  \end{split}
\end{equation}

\noindent where $\Phi_{\theta_0}^\prime$ is the restriction of $\Phi$ in $C(N, R_c)
\backslash C(N, R\theta_0)$. 

Once the node $\tau_0$ is determined, a second node $\tau_1$ must lie in the shadow 
region $A^+$ shown in Figures \ref{case2area1} or \ref{case3area1}, and a third node $\tau_2$ must lie in the 
region $S^-$ shown in Figures \ref{case3area1} or \ref{case2area2}, as illustrated in Section \ref{secboundcase2}.
The node $\tau_1 = (R, \theta_1, \varphi_1)$ is assumed to have the smallest
azimuth angle in $A^+$, which means that there should be no nodes with a azimuth angle
less than $\varphi_1$ in $A^+$, that is to say no nodes are in the region
\begin{displaymath}
  S^+(\tau_0, \tau_1) =  S^+(\alpha_0, \varphi_1) = A^+ \bigcap H^+(\varphi_1)
\end{displaymath}

Since the intensity measure of the Poisson point process in spherical
coordinates is $\lambda R^2 \sin \theta d\theta d\varphi$, the density $F_{\tau_1}$ of
$\tau_1$ can be given as
\begin{equation} \label{eqdens1}
  F_{\tau_1} (d\theta_1, d\varphi_1) = \lambda R^2 \sin \theta_1 e^{-\lambda |S^+(\alpha_0, \varphi_1)|} d\theta_1 d\varphi_1
\end{equation}

Then we derive the integration domain $D(\alpha_0)$ with respect to parameters $(\theta_1, \varphi_1)$. Consider the case shown in Figure \ref{case2area1},
assume the point $M_2$ has the spherical coordinate $(R, \alpha_0, \varphi_m), \varphi_m \in (0, \pi)$.
Since the great circle distance between $\tau_0$ and $M_2$ is $R_c$, 
then according to the spherical law of consines, we have
$\cos (R_c/R) = \cos^2 \alpha_0 + \sin ^2 \alpha_0 \cos \varphi_m 
\Rightarrow \varphi_m(\alpha_0) = \arccos [(\cos(R_c/R) - \cos^2 \alpha_0)/(\sin^2 \alpha_0)]$.
It can be seen that points $M_1$ and $Q$ have
the spherical coordinates $(R, \alpha_0, 2\pi - \varphi_m(\alpha_0))$ and
$(R, \alpha_0, 2\varphi_m(\alpha_0))$ respectively, where $Q$ is one intersection 
point between bases of spherical caps $C(N, R\alpha_0)$ and $C(M_2, R_c)$. Thus the integration
range for $\varphi_1$ is $[2\pi - \varphi_m(\alpha_0), 2\varphi_m(\alpha_0)]$.
In addition, assume any point with great circle distance $R_c$ to 
$\tau_0$ has the spherical coordinate $(R, \theta_t, \varphi_t)$,
still using the spherical law of consines, we have 
$\cos (R_c/R) = \cos \alpha_0 \cos \theta_t + \sin \alpha_0 \sin \theta_t \cos \varphi_t
\Rightarrow \theta_t(\alpha_0, \varphi_t) = \arccos [\cos(R_c/R)/\sqrt{1-\sin^2\alpha_0 \sin^2 \varphi_t}] + \arctan(\cos \varphi_t \tan \alpha_0)$.
Similarly, assume any point with great circle distance $R_c$
to $M_2$ has the spherical coordinate $(R, \theta_t', \varphi_t')$,
we can obtain $\theta_t'(\alpha_0, \varphi_t') = \arccos [\cos(R_c/R)/\sqrt{1-\sin^2\alpha_0 \sin^2 (\varphi_t' - \varphi_m(\alpha_0))}] + \arctan(\cos (\varphi_t' - \varphi_m(\alpha_0)) \tan \alpha_0)$. Then the integration range for $\theta_1$
is $[\alpha_0, \theta_{1u}(\alpha_0, \varphi_1)]$, where $\theta_{1u}(\alpha_0, \varphi_1) = \min \{\theta_{1u1}(\alpha_0, \varphi_1), \theta_{1u2}(\alpha_0, \varphi_1)\}$, $\theta_{1u1}(\alpha_0, \varphi_1) = \theta_t(\alpha_0, \varphi_1)$, $\theta_{1u2}(\alpha_0, \varphi_1) = \theta_t'(\alpha_0, \varphi_1)$.

Consider the case shown in Figure \ref{case3area1}, the derivation
of the integration domain $D(\alpha_0)$ is the same as the case shown 
in Figure \ref{case2area1}. In this case, the point $M$
has the spherical coordinate $(R, \alpha_0, \pi)$, and the integration range
for $\varphi_1$ is $[\pi, 2\pi]$. If we define
\[ \varphi_m(\alpha_0) = \left\{
  \begin{array}{l l}
    \pi & \quad \text{if $\frac{R_s}{R} < \alpha_0 \leq \frac{R_c}{2R}$}\\
    \arccos \frac{\cos\frac{R_c}{R} - \cos^2 \alpha_0}{\sin^2 \alpha_0} & \quad \text{othewise}
  \end{array} \right.\]
\noindent then the two cases can be regarded as the same in terms of the integration domain $D(\alpha_0)$.

Furthermore, $|S^+(\alpha_0, \varphi_1)|$ can be expressed as 
\begin{displaymath}
  |S^+(\alpha_0, \varphi_1)| = \int_{2\pi - \varphi_m(\alpha_0)}^{\varphi_1} \int_{\alpha_0}^{\theta_{1u}(\alpha_0, \varphi)} R^2 \sin \theta d\theta d\varphi
\end{displaymath}

As illustrated in Section \ref{secboundcase2}, assume only $\tau_0, \tau_1$ 
and nodes in $S^-(\tau_0, \tau_1)$ can
contribute to the spherical triangle which bounds a spherical triangular hole containing the
point $N$, we can get a lower bound of the probability that the
point $N$ is inside a spherical triangular hole. Based on the assumption,
we have
\begin{equation} \label{eqplower2}
  \begin{split}
    &\mathrm{P}\{\bigcup_{\{n_1, n_2\} \subseteq \Phi_{\theta_0}^\prime} T((R, \theta_0, 0), n_1, n_2)\}  \\
    & > \mathrm{P}\{\bigcup_{n_2 \subseteq \Phi_{\theta_0}^\prime \bigcap S^-(\tau_0, \tau_1) } T((R, \theta_0, 0), \tau_1, n_2)\}\\
    & = \iint\limits_{D(\theta_0)} \mathrm{P} \{ \bigcup_{n_2 \subseteq \Phi_{\theta_0}^\prime \bigcap \atop S^-(\theta_0, \theta_1, \varphi_1)} T((R, \theta_0, 0), (R, \theta_1, \varphi_1), n_2) \} \\
    & \qquad \qquad \qquad \qquad \qquad\qquad  F_{\tau_1} (d\theta_1, d\varphi_1) \\
    & = \iint_{D(\theta_0)} \mathrm{P} \{\Phi_{\theta_0}^\prime (S^-(\theta_0, \theta_1, \varphi_1)) > 0 \} F_{\tau_1} (d\theta_1, d\varphi_1)\\
    & = \iint_{D(\theta_0)} (1- e^{-\lambda |S^-(\theta_0, \theta_1, \varphi_1)|}) F_{\tau_1} (d\theta_1, d\varphi_1)
  \end{split}
\end{equation}

\noindent where $|S^-(\theta_0, \theta_1, \varphi_1)|$ can be expressed as
\begin{equation*} \label{eqareasminus}
 |S^-(\theta_0, \theta_1, \varphi_1)| = \int_{\varphi_{2l}}^{\varphi_m} \int_{\theta_0}^{\theta_{2u}} R^2 \sin \theta_2 d\theta_2 d\varphi_2
\end{equation*}

\noindent and
\begin{displaymath}
 \begin{split} 
   \varphi_{2l} &= \varphi_1 - \arccos \frac{\cos (R_c/R) - \cos \theta_1 \cos \theta_0}{\sin \theta_1 \sin \theta_0} \\
   \theta_{2u} &= \min \{\theta_{1u1}, \theta_{2u2} \} \\
   \theta_{2u2} &= \arccos \Big [\cos(R_c/R)/\sqrt{1-\sin^2\theta_0 \sin^2 (\varphi_2 - \varphi_1)} \Big] \\
   & \qquad + \arctan(\cos (\varphi_2 - \varphi_1) \tan \theta_1)
 \end{split}
\end{displaymath}

Therefore, from (\ref{eqdens0}), (\ref{eqplower1}), (\ref{eqdens1}) and (\ref{eqplower2}), 
the lower bound shown in (\ref{eqlower1}) can be derived.

As for the upper bound, replace $|S^-(\theta_0, \theta_1, \varphi_1)|$ by 
$|S^-(\theta_0, \theta_0, \varphi_1)|$, we can get the upper bound as illustrated in Section
 \ref{secboundcase2}.

\end{IEEEproof}

\section{Proof of Theorem \ref{theo2}} \label{app3}

Comparing (\ref{eqlower1}) to (\ref{eqlower2}), (\ref{equpper1}) to (\ref{equpper2}), we
can find that they are very similar. If we can show that each item related with $R$ in (\ref{eqlower1}) and (\ref{equpper1}) tends to its counterpart in (\ref{eqlower2}) and (\ref{equpper2}) when $R \to \infty$, then it is easy to prove Theorem \ref{theo2}. For convenience,  let $\theta_0 = r_0/R, \theta_1 = r_1/R, \varphi_1^\prime = \pi + \varphi_1$.

\begin{IEEEproof} First, we have 
\begin{equation} \label{eqappcond}
	\begin{split}
		& \lim_{R \to \infty} R \arccos([3\cos^2(R_s/R)-1]/2) \\
	  = & R_s \lim_{x \to 0} \frac{\arccos([3\cos^2(x)-1]/2)}{x} \quad (\textrm{let} \, x = R_s/R) \\
	  \overset{(a)}{=} & R_s \lim_{x \to 0} \frac{3 \cos x \sin x}{\sqrt{1-([3\cos^2(x)-1]/2)^2}} \\
	  = & R_s \lim_{x \to 0} \frac{6 \cos x \sin x}{\sqrt{(3-3\cos^2 x)(1+3\cos^2 x)}}
	  = \sqrt{3}R_s
	\end{split}	
\end{equation}
\noindent where $(a)$ follows from l'H\^{o}pital's rule.

From (\ref{eqappcond}), we know that when $R \to \infty$, the condition $R_c > R \arccos ([3\cos^2(R_s/R)-1]/2)$ is equivalent to the condition $R_c > \sqrt{3}R_s$.

Similarly, we can get 
\begin{equation} \label{eqappr0}
	\lim_{R \to \infty} R\theta_{0u} = \lim_{R \to \infty} R\arccos \sqrt{\frac{1+2\cos(R_c/R)}{3}} = \frac{R_c}{\sqrt{3}}
\end{equation}

Then, we can also obtain
\begin{equation} \label{eqappphim}
	\begin{split}
		& \lim_{R \to \infty} \arccos [(\cos(R_c/R) - \cos^2 \theta_0)/\sin^2 \theta_0] \\
	  = & \arccos (\lim_{R \to \infty} \frac{\cos(R_c/R) - \cos^2 (r_0/R)}{\sin^2 (r_0/R)} )\\
	  \overset{(a)}{=} & \arccos (\lim_{R \to \infty} \frac{\frac{R_c}{R^2} \sin \frac{R_c}{R} - \frac{2r_0}{R^2} \sin \frac{r_0}{R} \cos \frac{r_0}{R}}{-2r_0/R^2 \sin (r_0/R) \cos (r_0/R)}) \\
	  = & \arccos (1 - R_c^2/(2r_0^2))\\
	  \overset{(b)}{=} & \pi - 2\arccos(R_c/(2r_0))
	\end{split}
\end{equation}
\noindent where $(a)$ uses l'H\^{o}pital's rule and (b) follows from $\cos(\pi - 2\arccos(R_c/(2r_0))) = 1 - R_c^2/(2r_0^2)$ and $0 \leq \pi - 2\arccos(R_c/(2r_0)) \leq \pi$.

According to (\ref{eqappphim}), comparing (\ref{eqphim}) to (\ref{eqphil}) and (\ref{eqphiu}), we can get 
\begin{align}	
	 \varphi_l(r_0) & = \pi - \lim \limits_{R \to \infty} \varphi_m(\theta_0) \label{eqappphil}\\
	 \varphi_u(r_0) & = \lim \limits_{R \to \infty} 2\varphi_m(\theta_0) - \pi \label{eqappphiu}
\end{align}

Still using l'H\^{o}pital's rule, we can get the following results. The detailed calculation is omitted due to space limitation.
\begin{align}
	 \lim_{R \to \infty} R\theta_{1u1}(\theta_0, \varphi_1)& =
	\sqrt{R_c^2 - r_0^2\sin^2 \varphi_1^\prime} - r_0 \cos \varphi_1^\prime \label{eqappr1u1} \\
	\lim_{R \to \infty} R\theta_{1u2}(\theta_0, \varphi_1) & =
	 \sqrt{R_c^2 - r_0^2\sin^2 (\varphi_1^\prime+\varphi_l(r_0))} \label{eqappr1u2} \\
	 & \qquad + r_0 \cos (\varphi_1^\prime+\varphi_l(r_0)) \nonumber
\end{align}
\noindent where $\theta_{1u1}(\theta_0, \varphi_1)$ and $\theta_{1u2}(\theta_0, \varphi_1)$ are shown in (\ref{eqtheta1u1}) and (\ref{eqtheta1u2}).

From (\ref{eqappr1u1}) and (\ref{eqappr1u2}), comparing (\ref{eqtheta1u}) to (\ref{eqr1}), we have
\begin{equation} \label{eqappr1}
	 \lim_{R \to \infty} R\theta_{1u}(\theta_0, \varphi_1) 
 =  R_1(r_0, \varphi_1^\prime)
\end{equation}

From (\ref{eqappphil}), (\ref{eqappphiu}) and (\ref{eqappr1}), comparing (\ref{eqsplus}) to (\ref{eqsplusR}) and by some simple replacement, we can obtain
\begin{equation} \label{eqappsplus}
	\lim_{R \to \infty} |S^+(\theta_0, \varphi_1)| = |S^+(r_0, \varphi_1^\prime)|
\end{equation}

Similarly, we get
\begin{equation}  \label{eqappsminus}
	\lim_{R \to \infty} |S^-(\theta_0, \theta_1, \varphi_1)| = |S^-(r_0, r_1, \varphi_1^\prime)|
\end{equation}
\noindent where $|S^-(\theta_0, \theta_1, \varphi_1)|$ and $|S^-(r_0, r_1, \varphi_1^\prime)|$
are shown in (\ref{eqsminus}) and (\ref{eqsminusR}).

In addition, from (\ref{eqs}), we have 
\begin{equation} \label{eqapps0}
	\lim_{R \to \infty} |C(N, R\theta_0)| = \lim_{R \to \infty} 2\pi R^2 (1-\cos(r_0/R)) = \pi r_0^2
\end{equation}

Finally, using (\ref{eqappr0}), (\ref{eqappphil}), (\ref{eqappphiu}), (\ref{eqappr1}), (\ref{eqappsplus}), (\ref{eqappsminus}) and (\ref{eqapps0}), we can obtain from (\ref{eqlower1}) that
	\begin{align*}
		& \lim_{R \to \infty} p_l(\lambda) = \lim_{R \to \infty} 2\pi\lambda^2 \int_{R_s}^{R\theta_{0u}} R\sin \frac{r_0}{R} dr_0 \int_{\pi - \varphi_m(\theta_0)}^{2\varphi_m(\theta_0) - \pi} d\varphi_1^\prime\\ 
		& \qquad \qquad \int_{R\theta_0}^{R\theta_{1u}(\theta_0, \varphi_1)} R\sin \frac{r_1}{R}
		  \times e^{-\lambda \pi r_0^2} e^{-\lambda |S^+(r_0, \varphi_1^\prime)|} \\
		& \qquad \qquad \times (1 - e^{-\lambda |S^-(r_0, r_1, \varphi_1^\prime)|})dr_1 \\
		& = 2\pi\lambda ^2 \int_{R_s}^{R_c/\sqrt{3}} r_0 dr_0 \int_{\varphi_l(r_0)}^{\varphi_u(r_0)} d\varphi_1^\prime \int_{r_0}^{R_1(r_0, \varphi_1^\prime)}   e^{-\lambda \pi r_0^2}\\
		& \qquad \times e^{-\lambda |S^+(r_0, \varphi_1^\prime)|} (1 - e^{-\lambda |S^-(r_0, r_1, \varphi_1^\prime)|})r_1 dr_1 \\
		& = p_l^\prime(\lambda)
	\end{align*}
	Similarly, we can get $\lim\limits_{R \to \infty} p_u(\lambda) = p_u^\prime(\lambda)$.

\end{IEEEproof}



\ifCLASSOPTIONcaptionsoff
  \newpage
\fi



%

\bibliographystyle{IEEEtran}
\bibliography{IEEEabrv,TWC}

%

\begin{IEEEbiography}[{\includegraphics[width=1in,height=1.25in,clip,keepaspectratio]{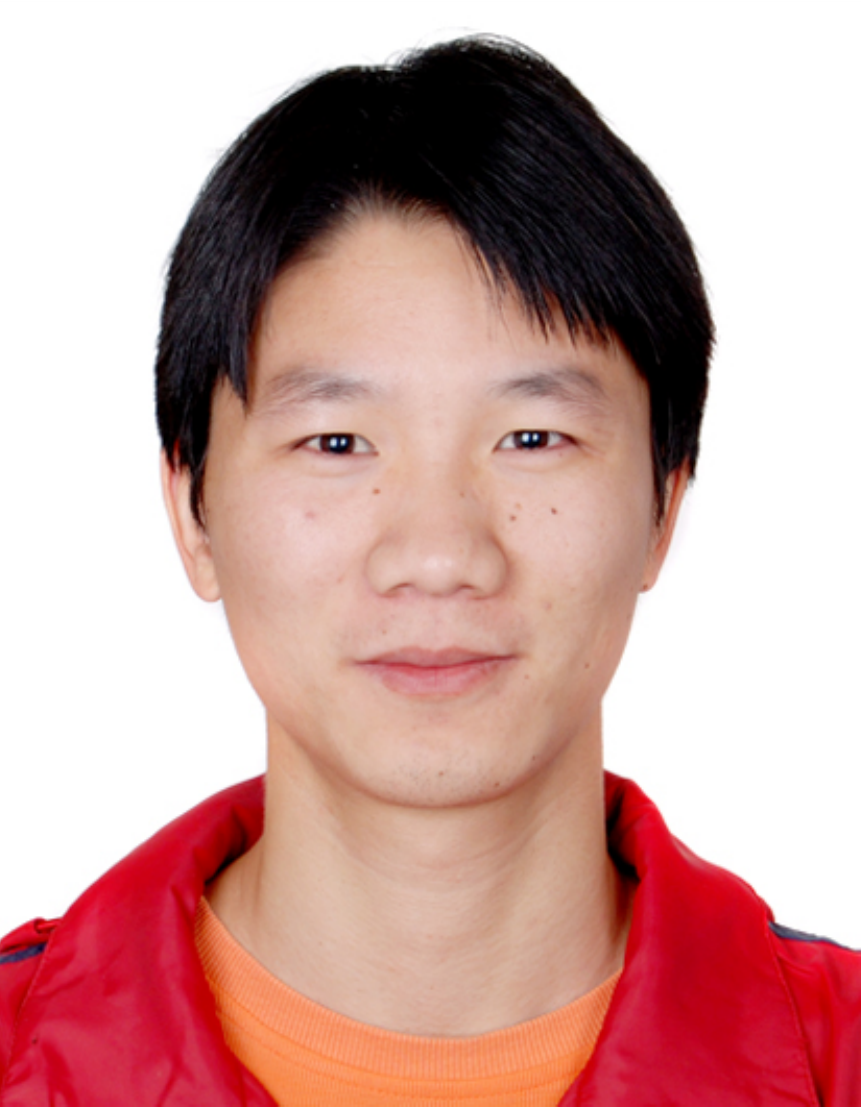}}]{Feng Yan}
received the B.S. degree from Huazhong University of Science and 
Technology, China, in 2005 and the M.S. degree from Southeast 
University, China, in 2008, both in electrical engineering. He 
received the Ph.D. degree in 2013 from Telecom ParisTech, Paris, France. He
is currently a postdoc in Telecom Bretagne, Rennes, France. His 
current research interests are in the areas of wireless 
communications and wireless networks, with emphasis on applications 
of homology theory, stochastic geometry in wireless networks.
\end{IEEEbiography}

\begin{IEEEbiography}[{\includegraphics[width=1in,height=1.25in,clip,keepaspectratio]{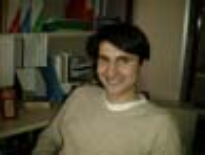}}]{Philippe Martins}
received a M.S. degree in signal processing and another M.S. degree 
in networking and computer science from Orsay University and ESIGETEL 
France, in 1996. He received the Ph.D. degree with honors in electrical 
engineering from Telecom ParisTech, Paris, France, in 2000.

He is currently a Professor in the Network and Computer Science
Department, at Telecom ParisTech. His main research interests lie in
performance evaluation in wireless networks (RRM, scheduling, handover
algorithms, radio metrology). His current investigations address mainly
three issues: a) the design of distributed sensing algorithms for
cognitive radio b) distributed coverage holes detection in wireless
sensor networks c) the definition of analytical models for the planning
and the dimensioning of cellular systems. He has published several papers 
on different international journals and
conferences. He is also an IEEE senior member and he is co-author of
several books on 3G and 4G systems.
\end{IEEEbiography}

\begin{IEEEbiography}[{\includegraphics[width=1in,height=1.25in,clip,keepaspectratio]{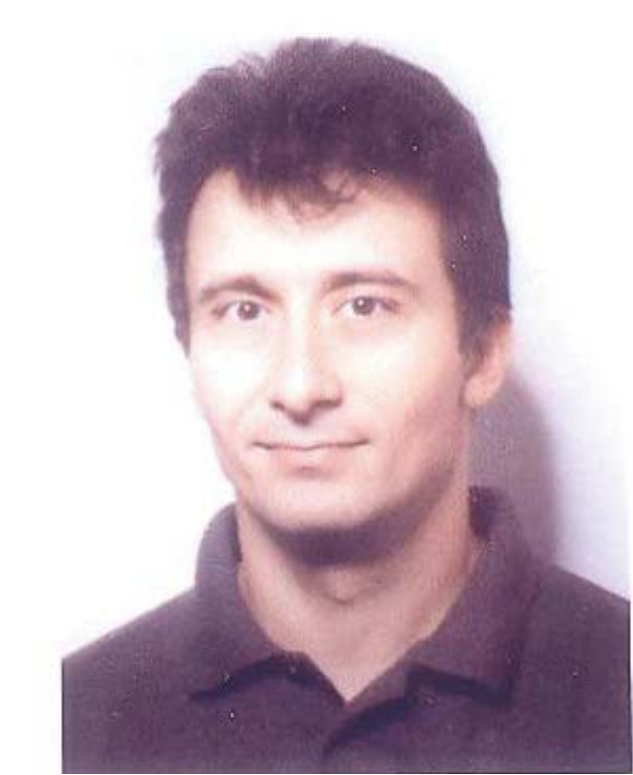}}]{Laurent Decreusefond}
is a former student of Ecole Normale Sup\'erieure de Cachan. He obtained his Ph.D. 
degree in Mathematics in 1994 from Telecom ParisTech and his Habilitation in 2001. 
He is currently a Professor in the Network and Computer Science
Department, at Telecom ParisTech. His main fields of interest are the Malliavin 
calculus, the stochastic analysis of long range dependent processes, random geometry 
and topology and their applications. With P. Moyal, he co-authored a book about 
the stochastic modelling of telecommunication systems.
\end{IEEEbiography}




\end{document}